\documentclass[article, aps, 12pt, floatfix, a4paper]{revtex4-2}

\usepackage{graphicx}
\usepackage{epstopdf}
\usepackage{amssymb}
\usepackage{hyperref}
\usepackage[normalem]{ulem}
\usepackage{color}
\usepackage{amsmath,bm}
\usepackage[utf8]{inputenc}
\usepackage{xcolor}
\usepackage{physics}
\usepackage{mathcomp}
\usepackage{multirow}
\usepackage{comment}

\usepackage{xr} 
\newcommand{\GRRA}{Gd$_3$(Ru$_{1-\delta}$Rh$_{\delta}$)$_4$Al$_{12}$}
\newcommand{\GRRAspec}{Gd$_3$(Ru$_{0.95}$Rh$_{0.05}$)$_4$Al$_{12}$}
\newcommand{\GRA}{Gd$_3$Ru$_4$Al$_{12}$}

\newcounter{hoge}
\makeatletter
\@addtoreset{figure}{hoge}
\makeatother
\makeatletter
\@addtoreset{equation}{hoge}
\makeatother
\makeatletter
\@addtoreset{section}{hoge}
\makeatother
\makeatletter
\@addtoreset{table}{hoge}
\makeatother

\begin{document}

\title[]{Metallic $p$-wave magnet with commensurate spin helix}


\author{Rinsuke Yamada$^{1,\P}$}\email{ryamada@ap.t.u-tokyo.ac.jp}
\author{Max T. Birch$^{2,\P}$}
\author{Priya R. Baral$^{1,\P}$}
\author{Shun Okumura$^{1}$}
\author{Ryota Nakano$^{1}$}
\author{Shang Gao$^{2,\S}$}
\author{Motohiko Ezawa$^{1}$}
\author{Takuya Nomoto$^{3}$}
\author{Jan Masell$^{2,4}$}
\author{Yuki Ishihara$^{1}$}
\author{Kamil K. Kolincio$^{2,5}$}
\author{Ilya Belopolski$^{2}$}
\author{Hajime Sagayama$^{6}$}
\author{Hironori Nakao$^{6}$}
\author{Kazuki Ohishi$^{7}$}
\author{Takashi Ohhara$^{8}$}
\author{Ryoji Kiyanagi$^{8}$}
\author{Taro Nakajima$^{9}$}
\author{Yoshinori Tokura$^{1,2,10}$}
\author{Taka-hisa Arima$^{2,11}$}
\author{Yukitoshi Motome$^{1}$}
\author{Moritz M. Hirschmann$^{2}$}
\author{Max Hirschberger$^{1,2}$}\email{hirschberger@ap.t.u-tokyo.ac.jp}

\affiliation{$^{1}$Department of Applied Physics and Quantum-Phase Electronics Center, The University of Tokyo, Bunkyo, Tokyo 113-8656, Japan}
\affiliation{$^{2}$RIKEN Center for Emergent Matter Science (CEMS), Wako, Saitama 351-0198, Japan}
\affiliation{$^{3}$Department of Physics, Tokyo Metropolitan University, Hachioji, Tokyo 192-0397, Japan}
\affiliation{$^{4}$Institute of Theoretical Solid State Physics, Karlsruhe Institute of Technology (KIT), 76049 Karlsruhe, Germany}
\affiliation{$^{5}$Faculty of Applied Physics and Mathematics, Gda{\'n}sk University of Technology, Narutowicza 11/12, 80-233 Gda{\'n}sk, Poland}
\affiliation{$^{6}$Institute of Materials Structure Science, High Energy Accelerator Research Organization, Tsukuba, Ibaraki 305-0801, Japan}
\affiliation{$^{7}$Neutron Science and Technology Center, Comprehensive Research Organization for Science and Society (CROSS), Tokai, Ibaraki 319-1106, Japan}
\affiliation{$^{8}$J-PARC Center, Japan Atomic Energy Agency, 2-4 Shirakata, Tokai, Ibaraki 319-1195 (Japan)}
\affiliation{$^{9}$The Institute for Solid State Physics, University of Tokyo, Kashiwa, Chiba 277-8561, Japan}
\affiliation{$^{10}$Tokyo College, University of Tokyo, Bunkyo, Tokyo 113-8656, Japan}
\affiliation{$^{11}$Department of Advanced Materials Science, The University of Tokyo, Kashiwa, Chiba 277-8561, Japan}
\affiliation{$^{\S}$Current address: Department of Physics, University of Science and Technology of China, Hefei 230026, China}
\affiliation{$^{\P}$Equal contribution}


\maketitle
\newpage


\begin{center}
\Large{Abstract}
\end{center}
\textbf{Antiferromagnetic states with spin-split electronic structure give rise to novel spintronic, magnonic, and electronic phenomena despite (near-) zero net magnetization~\cite{Ahn2019, Naka2019, Smejkal2020,Smejkal2022a,Smejkal2022b, Ezawa2024_spin_current, Yu2024}. \textcolor{black}{The simplest odd-parity spin splitting -- $p$-wave -- was originally proposed to emerge from a collective instability in interacting electron systems \cite{Hirsch1990, Wu2007, Jung2015, Kiselev2017, Wu2018}. Recent theory identifies a distinct route to realise $p$-wave spin-split electronic bands without strong correlations \cite{Hellenes2024,Jungwirth2024}, termed $p$-wave magnetism.
Here we demonstrate an experimental realisation of a metallic $p$-wave magnet.
The odd-parity spin splitting of delocalised conduction electrons arises from their coupling to an antiferromagnetic texture of localised magnetic moments: a coplanar spin helix whose magnetic period is an even multiple of the chemical unit cell, as revealed by X-ray scattering experiments.
This texture breaks space inversion symmetry but preserves time-reversal ($\mathcal{T}$) symmetry up to a half-unit-cell translation -- thereby fulfilling the symmetry conditions for $p$-wave magnetism.} Consistent with theoretical predictions, our $p$-wave magnet exhibits a characteristic anisotropy in the electronic conductivity~\cite{Hellenes2024,Jungwirth2024,Ezawa2024_electric_neel}. Relativistic spin-orbit coupling and a tiny spontaneous net magnetization further break $\mathcal{T}$ symmetry, resulting in a giant anomalous Hall effect (AHE, $\sigma_{xy}>600\,$S/cm, Hall angle $>3\,\%$), for an antiferromagnet. \textcolor{black}{Our model calculations show that the spin nodal planes found in the electronic structure of p-wave magnets are readily gapped by a small perturbation to induce the AHE.}}

\newpage
\newpage


\begin{center}
\Large{Main Text}
\end{center}

The spin splitting of electronic band states in momentum space is deeply intertwined with the \textcolor{black}{symmetries of magnetic order} in direct space~\cite{Naka2019, Ahn2019, Smejkal2020, Smejkal2022a,Smejkal2022b, Hellenes2024}. \textcolor{black}{Uniform ferromagnetic order} results in different energies for spin-up ($\uparrow$) and spin-down ($\downarrow$) bands, \textcolor{black}{producing a net spin polarisation} in momentum space \textcolor{black}{that is isotropic across the Fermi surface, as depicted} in Fig.~\ref{main:Fig1}\textcolor{black}{a}. In the language of spherical harmonics, \textcolor{black}{this corresponds to} an $s$-wave state \textcolor{black}{with even parity.} In contrast, $p$-wave \textcolor{black}{spin splitting is characterised in momentum space by} an anisotropic spin-split Fermi surface with \textcolor{black}{collinear $\uparrow$ and $\downarrow$ states exhibiting odd parity, shown in Fig.~\ref{main:Fig1}b.}

\textcolor{black}{In solids, $p$-wave spin splitting} has long been sought as the electronic analogue to the $A$-phase of superfluid helium-3. \textcolor{black}{An initial theoretical route to $p$-wave spin splitting was proposed in the form of a spin-channel Pomeranchuk instability — a spontaneous Fermi surface distortion driven by electronic correlations that breaks inversion symmetry but preserves the translational symmetry of the system (i.e., magnetic modulation vector $\bm{k}_\mathrm{mag}=0$)~\cite{Hirsch1990,Wu2007,Wu2018}.} \textcolor{black}{However, recent theory suggests that $p$-wave spin splitting also arises when materials form a magnetic superlattice with $\bm{k}_\mathrm{mag}\neq 0$, where the spin polarisation of electronic bands is now considered in the magnetic Brillouin zone~\cite{Hellenes2024,Jungwirth2024}.} 

\textcolor{black}{This scenario, termed $p$-wave magnet, can be realized without strong electronic correlations and arises when conduction electrons are coupled to an antiferromagnetic texture of localized magnetic moments obeying certain symmetries: Time reversal symmetry $\mathcal{T}$ paired with a half-unit cell translation $\bm{t}_{1/2}$ constrains the spin splitting to have odd parity or to vanish, while a $180$ degree spin rotation $C_{2\perp}$ combined with $\bm{t}_{1/2}$ ensures a unique axis for the spin expectation value of electronic states in momentum space -- say, having only $\left<S_x\right>$~\cite{Smejkal2022a,Smejkal2022b,Jungwirth2024}. To render $\left<S_x\right>$ finite, we require noncollinear magnetism and inversion symmetry breaking in practice, since collinear magnets only allow for even parity spin splitting~\cite{Smejkal2022b}.} \textcolor{black}{Strictly speaking, the} symmetry requirements \textcolor{black}{related to $\bm{t}_{1/2}$} are satisfied only by magnetic orders whose period is an even, integer multiple of the lattice constant: for example, the commensurate spin helix illustrated in Fig.~\ref{main:Fig1}b, \textcolor{black}{where the $p$-wave spin splitting is described by a spin polarisation vector $\bm{\alpha}$ (gray arrow) perpendicular to the rotation plane of magnetic moments (violet disk with looped arrow).}\\

\textcolor{black}{\textbf{Low-energy electronic structure of a $p$-wave magnet}}\\
\textcolor{black}{To show that a spin helix generically leads to an electronic structure with $p$-wave character, we use its symmetries $\left[\mathcal{T}\parallel\bm{t}_{1/2}\right]$ and $\left[C_{2\perp}\parallel\bm{t}_{1/2}\right]$ to constrain a single-orbital, low-energy model around $\bm{k}=0$.} \textcolor{black}{In the nonrelativistic limit, the only allowed terms are $\mathcal{H}_0(\bm{k}) = t_0\bm{k}^2 \sigma_{0} + p k_{x} \sigma_{x}$ where $\sigma_0$, $\sigma_x$, $t_0$ and $p$ are the identity matrix, a Pauli matrix, the transfer integral, and the magnitude of spin splitting, respectively (see Methods).} \textcolor{black}{This yields a spin-dependent band splitting with spin expectation value along $k_{x}$, described by the spin polarisation vector $\bm{\alpha}$ shown in Fig.~\ref{main:Fig1}c. Here, the spin-nodal plane is depicted as the line of states, indicated by the black parabola, on which $\uparrow$ (red) and $\downarrow$ (blue) bands touch.} Figure~\ref{main:Fig1}\textcolor{black}{d} shows the model's momentum-resolved expectation value for the spin component $\left<S_x\right>$ at the chemical potential. Anisotropic charge flow is naturally expected \textcolor{black}{in this model of a $p$-wave magnet} due to the difference in electronic structure along the $k_x$ and $k_y$ axes~\cite{Hellenes2024, Jungwirth2024, Brekke24, Ezawa2024_electric_neel,Ezawa2024_spin_current, Ezawa2024, Yu2024}. 

\textcolor{black}{The degeneracy of bands on the spin-nodal plane can be lifted by breaking of $\mathcal{T}$ symmetry using $m_z$ and explicitly} \textcolor{black}{ introducing a relativistic spin-orbit coupling $\lambda$ allowed by inversion symmetry breaking in the $p$-wave phase (Methods),}\textcolor{black}{
\begin{equation}
    \begin{split}
        \mathcal{H}(\bm{k}) = t_0\bm{k}^2 \sigma_{0} + p k_{x} \sigma_{x} + m_z \sigma_z + \lambda k_y \sigma_y.
    \end{split}
    \label{main:TwoBandPerturb}
\end{equation}
}
\textcolor{black}{The band hybridization induced by $m_z\neq 0$, $\lambda\neq 0$} \textcolor{black}{results in an anomalous Hall effect (AHE) due to the Berry curvature for the two bands shown in Fig.~\ref{main:Fig1}e, f respectively (see Methods).} \textcolor{black}{We stress that -- in contrast to other spin-split antiferromagnets of recent interest, such as $d$- or $g$-wave \textit{altermagnets}~\cite{Smejkal2020,Smejkal2022a,Smejkal2022b} -- the $p$-wave spin splitting does not, perforce, violate $\mathcal{T}$. Instead, the AHE emerges only in presence of a $\mathcal{T}$-breaking $m_z$.\\}

\textbf{Search for a candidate $p$-wave magnet}\\
\textcolor{black}{To realize a $p$-wave magnet based on an evenly commensurate spin helix \cite{Hellenes2024, Jungwirth2024}, we focus on the intermetallic compound \GRA{}, which is known to host coplanar spin helices which are incommensurate to the underlying atomic lattice~\cite{Gladyshevskii1993, Niermann2002, Nakamura2018, Matsumura2019}, and where the exchange interaction between the localised Gd moments is mediated by the conduction electrons. Replacing Rh for Ru and tuning the band filling in \GRRA{}, we shift the Fermi level -- and thus the period of the magnetic spin helix (see Methods and Supplementary Fig.~\ref{main:RhDopeDep}). This allows} us to approach an $N = 6$ commensurate spin helix, where both \textcolor{black}{$\left[\mathcal{T}\parallel\bm{t}_{1/2}\right]$} and  \textcolor{black}{$\left[C_{2\perp}\parallel\bm{t}_{1/2}\right]$} symmetries are satisfied. Around ${\delta}\approx 0.05$, \textcolor{black}{we identify a distinct magnetic ground state,} with signatures of $p$-wave magnetism, \textcolor{black}{as shown in the composition–temperature phase diagram of Fig.~\ref{main:Fig2}b.}

\textcolor{black}{To confirm that the magnetic order in the $\delta \approx 0.05$ phase is consistent with the symmetry requirements, we perform} resonant elastic X-ray scattering (REXS) measurements \textcolor{black}{on a series of single crystals with different Rh content $\delta$ (Methods). For each sample, we determine the magnetic} propagation vector $\bm{k}_\mathrm{mag}$ and its orientation relative to the crystal axes. The position of $\bm{k}_\mathrm{mag}$ in the Brillouin zone is shown schematically in the inset of Fig.~\ref{main:Fig2}b, where blue, red, and black symbols indicate $\bm{k}_\mathrm{mag}$ for $\textcolor{black}{\delta} = 0$, $0.05$, and $0.20$, respectively. In a \textcolor{black}{narrow composition window} around $\delta \approx 0.05$, \textcolor{black}{we find that the magnetic satellites are indexed as $(1/6, 1/6, 0)$ in Miller notation -- indicating that the magnetic period is evenly commensurate with the crystal lattice ($N = 6$). The reconstructed unit cell, spanning six crystallographic unit cells}, is outlined in orange in Fig.~\ref{main:Fig2}d. \textcolor{black}{Supporting neutron experiments are shown in Extended Data Fig.~\ref{main:taikan}.}\\

\textbf{Coplanar antiferromagnetism from resonant X-ray scattering}\\
\textcolor{black}{To satisfy {$\left[C_{2\perp}\parallel\bm{t}_{1/2}\right]$} symmetry, we specifically require coplanar magnetic order; that is, all magnetic moments should be perpendicular to a single axis.} \textcolor{black}{This is probed by} X-ray polarisation analysis measurements in the experimental geometry of Fig.~\ref{main:Fig2}c, where the incoming beam $\bm{k}_\mathrm{i}$ and the outgoing beam $\bm{k}_\mathrm{f}$ span the scattering plane (gray shaded, semi-circular plane\textcolor{black}{; see Methods}). The polarisation of incident X-rays is in the scattering plane ($\pi$-polarised), while we detect both horizontal ($\pi'$) and vertical ($\sigma'$) components of the scattered beam. The corresponding intensities are $I_{\pi - \pi'}$ and $I_{\pi - \sigma'}$, respectively. The $\pi - \pi'$ channel probes the magnetic moment perpendicular to the scattering plane, $I_{\pi- \pi'}\propto {\bm m}_{z}^2$, whereas the $\pi- \sigma'$ channel is $I_{\pi - \sigma'}\propto (\bm{k}_\mathrm{i} \cdot \bm{m}_{\mathrm{ip}})^2$~\cite{REXS_textbook} (see Methods). Here, $\bm{m}_{\mathrm{ip}}$ is the projection of the magnetic moment onto the scattering plane. \textcolor{black}{For each of the three magnetic reflections} illustrated in Fig.~\ref{main:Fig2}e -- \textcolor{black}{corresponding to three $p$-wave magnetic domains allowed by the hexagonal symmetry} -- we measured $I_{\pi - \pi'}$ and $I_{\pi - \sigma'}$, shown in Fig.~\ref{main:Fig2}f-h. The finite $I_{\pi - \pi'}$ intensity for all three reflections demonstrates that {$\bm{m}_{z}$ is nonzero. \textcolor{black}{In Fig. 2h, $I_{\pi - \sigma'}$ nearly vanishes, which we attribute to the relative angle between $\bm{k}_\mathrm{i}$ and $\bm{m}_{\mathrm{ip}}$ approaching 90°, consistent with a spin helix.} \textcolor{black}{These X-ray measurements identify the magnetic ground state as a coplanar spin helix, where every spin rotates within a common $\bm{m}_{z}$-$\bm{m}_\mathrm{ip}$ plane normal to the spin polarisation vector $\bm{\alpha}$ (shown in Fig.~\ref{main:Fig2}i).} This magnetic structure preserves the combined spin rotation and translation operation $[C_{2\perp} \parallel \bm{t}_{1/2}]$ as well as $[\mathcal{T}\parallel \bm{t}_{1/2}]$, \textcolor{black}{and therefore satisfies the symmetry requirements for $p$-wave magnetism.} \textcolor{black}{The spin polarisation vector $\bm{\alpha}$ for electronic states in momentum space, introduced in Fig.~\ref{main:Fig1}b, is then orthogonal to all local magnetic moments in direct space.}\\

\textbf{Electronic anisotropy of a $p$-wave magnet}\\
\textcolor{black}{One of the predicted consequences of the symmetry-protected $p$-wave state is in the electronic anisotropy,} where the directions parallel and perpendicular to the spin polarisation vector $\bm{\alpha}$ are not equivalent, \textcolor{black}{and we expect to see a resistance which depends on the angle of the applied current relative to the direction of $\bm{\alpha}$}~\cite{Hellenes2024, Jungwirth2024, Ezawa2024_electric_neel}. \textcolor{black}{To probe this, we utilised focused ion beam milling to fabricate} a freestanding, circular transport device \textcolor{black}{from a single crystal with $\delta = 0.05$.} \textcolor{black}{The meandering contacts relieve strain caused by thermal expansion mismatch with the substrate (Fig.~\ref{main:Fig3}a, Methods), helping to suppress extrinsic pinning of $\bm{\alpha}$.} \textcolor{black}{This device enables simultaneous measurement of the resistance} along three symmetry-equivalent directions of the hexagonal crystal structure, which become inequivalent \textcolor{black}{upon $p$-wave magnetic ordering (Fig.~\ref{main:Fig3}b)}. \textcolor{black}{Resistance $R_{xx}$ measured along the crystallographic $a$-axis as a function of temperature is shown in Fig.~\ref{main:Fig3}c, revealing} magnetic phase transitions \textcolor{black}{consistent with} the bulk magnetization (red).

Applying a finite magnetic field perpendicular to the $c$-axis and rotating the sample by the angle $\phi$, we observe a characteristic two-fold anisotropic resistivity in Fig.~\ref{main:Fig3}d-f below the magnetic transition -- \textcolor{black}{in contrast to the weak}, conventional anisotropic resistivity~\cite{McGuire1975} \textcolor{black}{observed} in the paramagnetic state (Fig.~\ref{main:Fig3}d, white circles). For each magnetic field direction, one of the three current directions $I_a$, $I_b$, or $I_c$ \textcolor{black}{exhibits a higher} resistance, \textcolor{black}{while} the other two have \textcolor{black}{lower} resistance. \textcolor{black}{This demonstrates that, for each sextant of the magnetic field angle $\phi$, one of the three magnetic domains characterised by distinct orientations of the spin polarisation vector $\bm{\alpha}$ is selected.} Following the conventions of Fig.~\ref{main:Fig1}a, the insets in Fig.~\ref{main:Fig3}d show the evolution of $\bm{\alpha}$ in discrete steps during the field rotation, and confirm that \textcolor{black}{current flows more easily} along the direction perpendicular to $\bm{\alpha}$. \textcolor{black}{In the present compound, the spin nodal plane is fixed at $k_x = 0$ by symmetry, so the two-fold anisotropic resistivity is directly related to the two-fold symmetry of the $p$-wave magnet's spin-split electronic structure (see Methods)}.\\

\textbf{Zero-field, giant anomalous Hall effect induced by time reversal breaking}\\
\textcolor{black}{Our $p$-wave magnet exhibits a tiny spontaneous magnetisation,} \textcolor{black}{resulting in a $\mathcal{T}$-breaking distortion of the $p$-wave electronic structure -- a scenario anticipated by models of spin helices of localized magnetic moments coupled to itinerant electrons~\cite{Okumura2018}.} \textcolor{black}{Returning to the bulk crystal with $\delta = 0.05$,} we set the magnetic field $B$ along the crystallographic $c$-axis and find that the magnetization increases roughly linearly and approaches the saturation value $M_\mathrm{sat}$ expected for Gd$^{3+}$ moments (Methods), as shown in Fig.~\ref{main:Fig4}a. Around zero magnetic field, we observe a small spontaneous magnetization $m_z$ of about $2\,\%$ of $M_\mathrm{sat}$ with a coercive field of about $0.15\,$T. \textcolor{black}{While the $p$-wave state is slightly distorted when $m_z$ becomes finite, the character of $p$-wave splitting -- or '$p$-waveness'~\cite{Hodt2024} -- remains robust in the regime of small $m_z$ (see Supplementary Fig.~\ref{main:pwaveness_multipanel}).} \textcolor{black}{The hysteretic behaviour is discussed in terms of weak exchange anisotropies in Supplementary Note~\ref{main:SI_ModelDoubleq}}.

\textcolor{black}{As outlined above, we argue that finite spin-orbit coupling in the real material, combined with slight $\mathcal{T}$-breaking of the $p$-wave bands in momentum space, can generate an intrinsic anomalous Hall effect driven by Berry curvature. Consistent with this expectation,} a \textcolor{black}{striking} transport response appears when $B$ is parallel to the $c$-axis: with the current in the $ab$ plane, a sharp, giant anomaly is observed in the Hall \textcolor{black}{conductivity $\sigma_{xy}$} around $B = 0$, which is smoothly suppressed \textcolor{black}{as $B$ approaches the critical field $B_\mathrm{c1}$.} \textcolor{black}{At higher magnetic field, in the incommensurate fan-type phase II and in the field-aligned ferromagnetic state, the Hall effect returns to conventional behaviour}, being composed of a magnetization-dependent term and a term linear in $B$: $\sigma_{xy} = S_{\color{black}{n}}B + S_{\color{black}m}\mu_{0}M$ with constants $S_{\color{black}n}$ and $S_{\color{black}m}$ (Methods, \textcolor{black}{Supplementary Fig.~\ref{main:PhaseII_SENJU}})~\cite{Nagaosa2010}. The strongly $B$-dependent contribution $\Delta \sigma_{xy}$ in the $p$-wave state is isolated by subtracting these two terms, and the resulting traces of $\Delta \sigma_{xy}$ are shown in Fig.~\ref{main:Fig4}c,d. \textcolor{black}{Returning to the minimal low energy model presented in Eq.~(\ref{main:TwoBandPerturb}), we can discuss the experimentally observed $B$-field dependence of the AHE. The result is shown in} Figure~\ref{main:Fig4}f, \textcolor{black}{which demonstrates} a sharp anomaly in the anomalous Hall conductivity $\sigma_{xy}$ around zero magnetic field \textcolor{black}{or around zero $m_z$}, when $\left|\lambda\right| < \left|p\right|$ \textcolor{black}{(see Supplementary Note~\ref{main:conversion_exchange_splitting})}. \textcolor{black}{Finally, Fig.}~\ref{main:Fig4}e compares the zero-field AHE in various bulk antiferromagnets, highlighting the large $\Delta \sigma_{xy}$ response in our $p$-wave candidate, which exceeds $600\,$S/cm -- comparable to \textcolor{black}{the intrinsic AHE of} ferromagnets. Unlike other antiferromagnets with large AHE and small net magnetization~\cite{Nakatsuji2015, Nayak2016, Ghimire2018, Takagi2023, Park2023, Smejkal2022NRM}, $M(H)$ here saturates at modest, accessible fields, enabling practical switching between the $p$-wave and \textcolor{black}{field-aligned} ferromagnetic states. \\

\textbf{Conclusions}\\
The \textcolor{black}{experimentally observed,} commensurate spin helix with $N = 6$ period in direct space satisfies the symmetry requirements \textcolor{black}{for a $p$-wave magnet} up to a tiny, uniform spin splitting $m_z$.
\textcolor{black}{This $p$-wave magnet exhibits anisotropic electronic properties and} a surprisingly large anomalous Hall effect in zero magnetic field. \textcolor{black}{The Hall response is attributed to Berry curvature in momentum space arising when $m_z$ opens a gap at the nodal planes of a $p$-wave magnet.} Recent theoretical studies predict the generation of large linear and non-linear spin currents in $p$-wave magnets as well as a large tunneling magnetoresistance~\cite{Brekke2024, Hedayati2024,Ezawa2024_electric_neel,Ezawa2024_spin_current, Pari2024, Yu2024}. In particular, it is believed that the low number of nodal planes in spin-split $p$-wave antiferromagnets is advantageous for functionality as a spin injector~\cite{Brekke2024, Hedayati2024}. \textcolor{black}{Expanding on the literature of proximitized spin helices as a route to unconventional superconductivity~\cite{Choy2011, Martin2012, NadjPerge2013, Klinovaja2013}, }\textcolor{black}{superconducting proximity effects have also been recently discussed for $p$-wave magnets~\cite{Maeda2024,Ezawa2024}.} \textcolor{black}{Our findings establish metallic} $p$-wave magnets as an ideal platform to explore the interplay of \textcolor{black}{spin-split electronic states in magnets}, spintronics, superconductivity and electronic band topology. \\

\textcolor{black}{Note added: We became aware of the realization of an insulating odd-parity wave magnet after submission of this work~\cite{Song2025}.}


\newpage

\bibliography{main_only}

\begin{thebibliography}{55}%
\makeatletter
\providecommand \@ifxundefined [1]{%
 \@ifx{#1\undefined}
}%
\providecommand \@ifnum [1]{%
 \ifnum #1\expandafter \@firstoftwo
 \else \expandafter \@secondoftwo
 \fi
}%
\providecommand \@ifx [1]{%
 \ifx #1\expandafter \@firstoftwo
 \else \expandafter \@secondoftwo
 \fi
}%
\providecommand \natexlab [1]{#1}%
\providecommand \enquote  [1]{``#1''}%
\providecommand \bibnamefont  [1]{#1}%
\providecommand \bibfnamefont [1]{#1}%
\providecommand \citenamefont [1]{#1}%
\providecommand \href@noop [0]{\@secondoftwo}%
\providecommand \href [0]{\begingroup \@sanitize@url \@href}%
\providecommand \@href[1]{\@@startlink{#1}\@@href}%
\providecommand \@@href[1]{\endgroup#1\@@endlink}%
\providecommand \@sanitize@url [0]{\catcode `\\12\catcode `\$12\catcode `\&12\catcode `\#12\catcode `\^12\catcode `\_12\catcode `\%12\relax}%
\providecommand \@@startlink[1]{}%
\providecommand \@@endlink[0]{}%
\providecommand \url  [0]{\begingroup\@sanitize@url \@url }%
\providecommand \@url [1]{\endgroup\@href {#1}{\urlprefix }}%
\providecommand \urlprefix  [0]{URL }%
\providecommand \Eprint [0]{\href }%
\providecommand \doibase [0]{https://doi.org/}%
\providecommand \selectlanguage [0]{\@gobble}%
\providecommand \bibinfo  [0]{\@secondoftwo}%
\providecommand \bibfield  [0]{\@secondoftwo}%
\providecommand \translation [1]{[#1]}%
\providecommand \BibitemOpen [0]{}%
\providecommand \bibitemStop [0]{}%
\providecommand \bibitemNoStop [0]{.\EOS\space}%
\providecommand \EOS [0]{\spacefactor3000\relax}%
\providecommand \BibitemShut  [1]{\csname bibitem#1\endcsname}%
\let\auto@bib@innerbib\@empty
\bibitem [{\citenamefont {Ahn}\ \emph {et~al.}(2019)\citenamefont {Ahn}, \citenamefont {Hariki}, \citenamefont {Lee},\ and\ \citenamefont {Kune\ifmmode~\check{s}\else \v{s}\fi{}}}]{Ahn2019}%
  \BibitemOpen
  \bibfield  {author} {\bibinfo {author} {\bibfnamefont {K.-H.}\ \bibnamefont {Ahn}}, \bibinfo {author} {\bibfnamefont {A.}~\bibnamefont {Hariki}}, \bibinfo {author} {\bibfnamefont {K.-W.}\ \bibnamefont {Lee}},\ and\ \bibinfo {author} {\bibfnamefont {J.}~\bibnamefont {Kune\ifmmode~\check{s}\else \v{s}\fi{}}},\ }\bibfield  {title} {\bibinfo {title} {{Antiferromagnetism in ${\mathrm{RuO}}_{2}$ as $d$-wave Pomeranchuk instability}},\ }\href@noop {} {\bibfield  {journal} {\bibinfo  {journal} {Phys. Rev. B}\ }\textbf {\bibinfo {volume} {99}},\ \bibinfo {pages} {184432} (\bibinfo {year} {2019})}\BibitemShut {NoStop}%
\bibitem [{\citenamefont {Naka}\ \emph {et~al.}(2019)\citenamefont {Naka}, \citenamefont {Hayami}, \citenamefont {Kusunose}, \citenamefont {Yanagi}, \citenamefont {Motome},\ and\ \citenamefont {Seo}}]{Naka2019}%
  \BibitemOpen
  \bibfield  {author} {\bibinfo {author} {\bibfnamefont {M.}~\bibnamefont {Naka}}, \bibinfo {author} {\bibfnamefont {S.}~\bibnamefont {Hayami}}, \bibinfo {author} {\bibfnamefont {H.}~\bibnamefont {Kusunose}}, \bibinfo {author} {\bibfnamefont {Y.}~\bibnamefont {Yanagi}}, \bibinfo {author} {\bibfnamefont {Y.}~\bibnamefont {Motome}},\ and\ \bibinfo {author} {\bibfnamefont {H.}~\bibnamefont {Seo}},\ }\bibfield  {title} {\bibinfo {title} {{Spin current generation in organic antiferromagnets}},\ }\href@noop {} {\bibfield  {journal} {\bibinfo  {journal} {Nat. Commun.}\ }\textbf {\bibinfo {volume} {10}},\ \bibinfo {pages} {4305} (\bibinfo {year} {2019})}\BibitemShut {NoStop}%
\bibitem [{\citenamefont {\ifmmode~\check{S}\else \v{S}\fi{}mejkal}\ \emph {et~al.}(2020)\citenamefont {\ifmmode~\check{S}\else \v{S}\fi{}mejkal}, \citenamefont {Gonz\'{a}lez-Hern\'{a}ndez}, \citenamefont {Jungwirth},\ and\ \citenamefont {Sinova}}]{Smejkal2020}%
  \BibitemOpen
  \bibfield  {author} {\bibinfo {author} {\bibfnamefont {L.}~\bibnamefont {\ifmmode~\check{S}\else \v{S}\fi{}mejkal}}, \bibinfo {author} {\bibfnamefont {R.}~\bibnamefont {Gonz\'{a}lez-Hern\'{a}ndez}}, \bibinfo {author} {\bibfnamefont {T.}~\bibnamefont {Jungwirth}},\ and\ \bibinfo {author} {\bibfnamefont {J.}~\bibnamefont {Sinova}},\ }\bibfield  {title} {\bibinfo {title} {{Crystal time-reversal symmetry breaking and spontaneous Hall effect in collinear antiferromagnets}},\ }\href@noop {} {\bibfield  {journal} {\bibinfo  {journal} {Sci. Adv.}\ }\textbf {\bibinfo {volume} {6}},\ \bibinfo {pages} {eaaz8809} (\bibinfo {year} {2020})}\BibitemShut {NoStop}%
\bibitem [{\citenamefont {\ifmmode~\check{S}\else \v{S}\fi{}mejkal}\ \emph {et~al.}(2022{\natexlab{a}})\citenamefont {\ifmmode~\check{S}\else \v{S}\fi{}mejkal}, \citenamefont {Sinova},\ and\ \citenamefont {Jungwirth}}]{Smejkal2022a}%
  \BibitemOpen
  \bibfield  {author} {\bibinfo {author} {\bibfnamefont {L.}~\bibnamefont {\ifmmode~\check{S}\else \v{S}\fi{}mejkal}}, \bibinfo {author} {\bibfnamefont {J.}~\bibnamefont {Sinova}},\ and\ \bibinfo {author} {\bibfnamefont {T.}~\bibnamefont {Jungwirth}},\ }\bibfield  {title} {\bibinfo {title} {{Emerging research landscape of altermagnetism}},\ }\href@noop {} {\bibfield  {journal} {\bibinfo  {journal} {Phys. Rev. X}\ }\textbf {\bibinfo {volume} {12}},\ \bibinfo {pages} {040501} (\bibinfo {year} {2022}{\natexlab{a}})}\BibitemShut {NoStop}%
\bibitem [{\citenamefont {\ifmmode~\check{S}\else \v{S}\fi{}mejkal}\ \emph {et~al.}(2022{\natexlab{b}})\citenamefont {\ifmmode~\check{S}\else \v{S}\fi{}mejkal}, \citenamefont {Sinova},\ and\ \citenamefont {Jungwirth}}]{Smejkal2022b}%
  \BibitemOpen
  \bibfield  {author} {\bibinfo {author} {\bibfnamefont {L.}~\bibnamefont {\ifmmode~\check{S}\else \v{S}\fi{}mejkal}}, \bibinfo {author} {\bibfnamefont {J.}~\bibnamefont {Sinova}},\ and\ \bibinfo {author} {\bibfnamefont {T.}~\bibnamefont {Jungwirth}},\ }\bibfield  {title} {\bibinfo {title} {{Beyond Conventional Ferromagnetism and Antiferromagnetism: A Phase with Nonrelativistic Spin and Crystal Rotation Symmetry}},\ }\href@noop {} {\bibfield  {journal} {\bibinfo  {journal} {Phys. Rev. X}\ }\textbf {\bibinfo {volume} {12}},\ \bibinfo {pages} {031042} (\bibinfo {year} {2022}{\natexlab{b}})}\BibitemShut {NoStop}%
\bibitem [{\citenamefont {Ezawa}(2024{\natexlab{a}})}]{Ezawa2024_spin_current}%
  \BibitemOpen
  \bibfield  {author} {\bibinfo {author} {\bibfnamefont {M.}~\bibnamefont {Ezawa}},\ }\bibfield  {title} {\bibinfo {title} {{Third-order and fifth-order nonlinear spin-current generation in $g$-wave and $i$-wave altermagnets and perfect spin-current diode based on $f$-wave magnets}},\ }\href@noop {} {\bibfield  {journal} {\bibinfo  {journal} {arXiv:2411.16036}\ } (\bibinfo {year} {2024}{\natexlab{a}})}\BibitemShut {NoStop}%
\bibitem [{\citenamefont {Yu}\ \emph {et~al.}(2024)\citenamefont {Yu}, \citenamefont {Lyngby}, \citenamefont {Shishidou}, \citenamefont {Roig}, \citenamefont {Kreisel}, \citenamefont {Weinert}, \citenamefont {Andersen}, ,\ and\ \citenamefont {Agterberg}}]{Yu2024}%
  \BibitemOpen
  \bibfield  {author} {\bibinfo {author} {\bibfnamefont {Y.}~\bibnamefont {Yu}}, \bibinfo {author} {\bibfnamefont {M.~B.}\ \bibnamefont {Lyngby}}, \bibinfo {author} {\bibfnamefont {T.}~\bibnamefont {Shishidou}}, \bibinfo {author} {\bibfnamefont {M.}~\bibnamefont {Roig}}, \bibinfo {author} {\bibfnamefont {A.}~\bibnamefont {Kreisel}}, \bibinfo {author} {\bibfnamefont {M.}~\bibnamefont {Weinert}}, \bibinfo {author} {\bibfnamefont {B.~M.}\ \bibnamefont {Andersen}}, ,\ and\ \bibinfo {author} {\bibfnamefont {D.~F.}\ \bibnamefont {Agterberg}},\ }\bibfield  {title} {\bibinfo {title} {{Odd-parity magnetism driven by antiferromagnetic exchange}},\ }\href@noop {} {\bibfield  {journal} {\bibinfo  {journal} {arXiv preprint arXiv:2501.02057v2}\ } (\bibinfo {year} {2024})}\BibitemShut {NoStop}%
\bibitem [{\citenamefont {Hirsch}(1990)}]{Hirsch1990}%
  \BibitemOpen
  \bibfield  {author} {\bibinfo {author} {\bibfnamefont {J.~E.}\ \bibnamefont {Hirsch}},\ }\bibfield  {title} {\bibinfo {title} {{Spin-split states in metals}},\ }\href@noop {} {\bibfield  {journal} {\bibinfo  {journal} {Phys. Rev. B}\ }\textbf {\bibinfo {volume} {41}},\ \bibinfo {pages} {6820} (\bibinfo {year} {1990})}\BibitemShut {NoStop}%
\bibitem [{\citenamefont {Wu}\ \emph {et~al.}(2007)\citenamefont {Wu}, \citenamefont {Sun}, \citenamefont {Fradkin},\ and\ \citenamefont {Zhang}}]{Wu2007}%
  \BibitemOpen
  \bibfield  {author} {\bibinfo {author} {\bibfnamefont {C.}~\bibnamefont {Wu}}, \bibinfo {author} {\bibfnamefont {K.}~\bibnamefont {Sun}}, \bibinfo {author} {\bibfnamefont {E.}~\bibnamefont {Fradkin}},\ and\ \bibinfo {author} {\bibfnamefont {S.-C.}\ \bibnamefont {Zhang}},\ }\bibfield  {title} {\bibinfo {title} {{Fermi liquid instabilities in the spin channel}},\ }\href@noop {} {\bibfield  {journal} {\bibinfo  {journal} {Phys. Rev. B}\ }\textbf {\bibinfo {volume} {75}},\ \bibinfo {pages} {115103} (\bibinfo {year} {2007})}\BibitemShut {NoStop}%
\bibitem [{\citenamefont {Jung}\ \emph {et~al.}(2015)\citenamefont {Jung}, \citenamefont {Polini},\ and\ \citenamefont {MacDonald}}]{Jung2015}%
  \BibitemOpen
  \bibfield  {author} {\bibinfo {author} {\bibfnamefont {J.}~\bibnamefont {Jung}}, \bibinfo {author} {\bibfnamefont {M.}~\bibnamefont {Polini}},\ and\ \bibinfo {author} {\bibfnamefont {A.~H.}\ \bibnamefont {MacDonald}},\ }\bibfield  {title} {\bibinfo {title} {{Persistent current states in bilayer graphene}},\ }\href@noop {} {\bibfield  {journal} {\bibinfo  {journal} {Phys. Rev. B}\ }\textbf {\bibinfo {volume} {91}},\ \bibinfo {pages} {155423} (\bibinfo {year} {2015})}\BibitemShut {NoStop}%
\bibitem [{\citenamefont {Kiselev}\ \emph {et~al.}(2017)\citenamefont {Kiselev}, \citenamefont {Scheurer}, \citenamefont {W\"olfle},\ and\ \citenamefont {Schmalian}}]{Kiselev2017}%
  \BibitemOpen
  \bibfield  {author} {\bibinfo {author} {\bibfnamefont {E.~I.}\ \bibnamefont {Kiselev}}, \bibinfo {author} {\bibfnamefont {M.~S.}\ \bibnamefont {Scheurer}}, \bibinfo {author} {\bibfnamefont {P.}~\bibnamefont {W\"olfle}},\ and\ \bibinfo {author} {\bibfnamefont {J.}~\bibnamefont {Schmalian}},\ }\bibfield  {title} {\bibinfo {title} {{Limits on dynamically generated spin-orbit coupling: Absence of $l=1$ Pomeranchuk instabilities in metals}},\ }\href@noop {} {\bibfield  {journal} {\bibinfo  {journal} {Phys. Rev. B}\ }\textbf {\bibinfo {volume} {95}},\ \bibinfo {pages} {125122} (\bibinfo {year} {2017})}\BibitemShut {NoStop}%
\bibitem [{\citenamefont {Wu}\ \emph {et~al.}(2018)\citenamefont {Wu}, \citenamefont {Klein},\ and\ \citenamefont {Chubukov}}]{Wu2018}%
  \BibitemOpen
  \bibfield  {author} {\bibinfo {author} {\bibfnamefont {Y.-M.}\ \bibnamefont {Wu}}, \bibinfo {author} {\bibfnamefont {A.}~\bibnamefont {Klein}},\ and\ \bibinfo {author} {\bibfnamefont {A.~V.}\ \bibnamefont {Chubukov}},\ }\bibfield  {title} {\bibinfo {title} {{Conditions for $l=1$ Pomeranchuk instability in a Fermi liquid}},\ }\href@noop {} {\bibfield  {journal} {\bibinfo  {journal} {Phys. Rev. B}\ }\textbf {\bibinfo {volume} {97}},\ \bibinfo {pages} {165101} (\bibinfo {year} {2018})}\BibitemShut {NoStop}%
\bibitem [{\citenamefont {Hellenes}\ \emph {et~al.}(2024)\citenamefont {Hellenes}, \citenamefont {Jungwirth}, \citenamefont {Rodrigo Jaeschke-Ubiergo}, \citenamefont {Sinova},\ and\ \citenamefont {\ifmmode~\check{S}\else \v{S}\fi{}mejkal}}]{Hellenes2024}%
  \BibitemOpen
  \bibfield  {author} {\bibinfo {author} {\bibfnamefont {A.~B.}\ \bibnamefont {Hellenes}}, \bibinfo {author} {\bibfnamefont {T.}~\bibnamefont {Jungwirth}}, \bibinfo {author} {\bibfnamefont {A.~C.}\ \bibnamefont {Rodrigo Jaeschke-Ubiergo}}, \bibinfo {author} {\bibfnamefont {J.}~\bibnamefont {Sinova}},\ and\ \bibinfo {author} {\bibfnamefont {L.}~\bibnamefont {\ifmmode~\check{S}\else \v{S}\fi{}mejkal}},\ }\bibfield  {title} {\bibinfo {title} {{P-wave magnets}},\ }\href@noop {} {\bibfield  {journal} {\bibinfo  {journal} {arXiv preprint arXiv:2309.01607}\ } (\bibinfo {year} {2024})}\BibitemShut {NoStop}%
\bibitem [{\citenamefont {Jungwirth}\ \emph {et~al.}(2024)\citenamefont {Jungwirth}, \citenamefont {Fernandes}, \citenamefont {Fradkin}, \citenamefont {MacDonald}, \citenamefont {Sinova},\ and\ \citenamefont {Smejkal}}]{Jungwirth2024}%
  \BibitemOpen
  \bibfield  {author} {\bibinfo {author} {\bibfnamefont {T.}~\bibnamefont {Jungwirth}}, \bibinfo {author} {\bibfnamefont {R.~M.}\ \bibnamefont {Fernandes}}, \bibinfo {author} {\bibfnamefont {E.}~\bibnamefont {Fradkin}}, \bibinfo {author} {\bibfnamefont {A.~H.}\ \bibnamefont {MacDonald}}, \bibinfo {author} {\bibfnamefont {J.}~\bibnamefont {Sinova}},\ and\ \bibinfo {author} {\bibfnamefont {L.}~\bibnamefont {Smejkal}},\ }\bibfield  {title} {\bibinfo {title} {{From superfluid $^3$He to altermagnets}},\ }\href@noop {} {\bibfield  {journal} {\bibinfo  {journal} {arXiv:2411.00717}\ } (\bibinfo {year} {2024})}\BibitemShut {NoStop}%
\bibitem [{\citenamefont {Ezawa}(2024{\natexlab{b}})}]{Ezawa2024_electric_neel}%
  \BibitemOpen
  \bibfield  {author} {\bibinfo {author} {\bibfnamefont {M.}~\bibnamefont {Ezawa}},\ }\bibfield  {title} {\bibinfo {title} {{Purely electrical detection of the N{\'e}el vector of $p$-wave magnets based on linear and nonlinear conductivities}},\ }\href@noop {} {\bibfield  {journal} {\bibinfo  {journal} {arXiv:2410.21854}\ } (\bibinfo {year} {2024}{\natexlab{b}})}\BibitemShut {NoStop}%
\bibitem [{\citenamefont {Brekke}\ \emph {et~al.}(2024{\natexlab{a}})\citenamefont {Brekke}, \citenamefont {Sukhachov}, \citenamefont {Giil}, \citenamefont {Brataas},\ and\ \citenamefont {Linder}}]{Brekke24}%
  \BibitemOpen
  \bibfield  {author} {\bibinfo {author} {\bibfnamefont {B.}~\bibnamefont {Brekke}}, \bibinfo {author} {\bibfnamefont {P.}~\bibnamefont {Sukhachov}}, \bibinfo {author} {\bibfnamefont {H.~G.}\ \bibnamefont {Giil}}, \bibinfo {author} {\bibfnamefont {A.}~\bibnamefont {Brataas}},\ and\ \bibinfo {author} {\bibfnamefont {J.}~\bibnamefont {Linder}},\ }\bibfield  {title} {\bibinfo {title} {{Minimal Models and Transport Properties of Unconventional $p$-Wave Magnets}},\ }\href@noop {} {\bibfield  {journal} {\bibinfo  {journal} {Phys. Rev. Lett.}\ }\textbf {\bibinfo {volume} {133}},\ \bibinfo {pages} {236703} (\bibinfo {year} {2024}{\natexlab{a}})}\BibitemShut {NoStop}%
\bibitem [{\citenamefont {Ezawa}(2024{\natexlab{c}})}]{Ezawa2024}%
  \BibitemOpen
  \bibfield  {author} {\bibinfo {author} {\bibfnamefont {M.}~\bibnamefont {Ezawa}},\ }\bibfield  {title} {\bibinfo {title} {{Topological insulators and superconductors based on $p$-wave magnets: Electrical control and detection of a domain wall}},\ }\href@noop {} {\bibfield  {journal} {\bibinfo  {journal} {Phys. Rev. B}\ }\textbf {\bibinfo {volume} {110}},\ \bibinfo {pages} {165429} (\bibinfo {year} {2024}{\natexlab{c}})}\BibitemShut {NoStop}%
\bibitem [{\citenamefont {Gladyshevskii}\ \emph {et~al.}(1993)\citenamefont {Gladyshevskii}, \citenamefont {Strusievicz}, \citenamefont {Cenzual},\ and\ \citenamefont {Parthé}}]{Gladyshevskii1993}%
  \BibitemOpen
  \bibfield  {author} {\bibinfo {author} {\bibfnamefont {R.~E.}\ \bibnamefont {Gladyshevskii}}, \bibinfo {author} {\bibfnamefont {O.~R.}\ \bibnamefont {Strusievicz}}, \bibinfo {author} {\bibfnamefont {K.}~\bibnamefont {Cenzual}},\ and\ \bibinfo {author} {\bibfnamefont {E.}~\bibnamefont {Parthé}},\ }\bibfield  {title} {\bibinfo {title} {{Structure of Gd$_3$Ru$_4$Al$_{12}$, a new member of the EuMg$_{5.2}$ structure family with minority-atom clusters}},\ }\href@noop {} {\bibfield  {journal} {\bibinfo  {journal} {Acta Crystallogr. B}\ }\textbf {\bibinfo {volume} {49}},\ \bibinfo {pages} {474} (\bibinfo {year} {1993})}\BibitemShut {NoStop}%
\bibitem [{\citenamefont {Niermann}\ and\ \citenamefont {Jeitschko}(2002)}]{Niermann2002}%
  \BibitemOpen
  \bibfield  {author} {\bibinfo {author} {\bibfnamefont {J.}~\bibnamefont {Niermann}}\ and\ \bibinfo {author} {\bibfnamefont {W.}~\bibnamefont {Jeitschko}},\ }\bibfield  {title} {\bibinfo {title} {{Ternary rare earth ($R$) transition metal aluminides $R_{3}T_{4}$Al$_{12}$ ($T$ = Ru and Os) with Gd$_3$Ru$_4$Al$_{12}$ type structure}},\ }\href@noop {} {\bibfield  {journal} {\bibinfo  {journal} {Z. Anorg. Allg. Chem.}\ }\textbf {\bibinfo {volume} {628}},\ \bibinfo {pages} {2549} (\bibinfo {year} {2002})}\BibitemShut {NoStop}%
\bibitem [{\citenamefont {Nakamura}\ \emph {et~al.}(2018)\citenamefont {Nakamura}, \citenamefont {Kabeya}, \citenamefont {Kobayashi}, \citenamefont {Araki}, \citenamefont {Katoh},\ and\ \citenamefont {Ochiai}}]{Nakamura2018}%
  \BibitemOpen
  \bibfield  {author} {\bibinfo {author} {\bibfnamefont {S.}~\bibnamefont {Nakamura}}, \bibinfo {author} {\bibfnamefont {N.}~\bibnamefont {Kabeya}}, \bibinfo {author} {\bibfnamefont {M.}~\bibnamefont {Kobayashi}}, \bibinfo {author} {\bibfnamefont {K.}~\bibnamefont {Araki}}, \bibinfo {author} {\bibfnamefont {K.}~\bibnamefont {Katoh}},\ and\ \bibinfo {author} {\bibfnamefont {A.}~\bibnamefont {Ochiai}},\ }\bibfield  {title} {\bibinfo {title} {{Spin trimer formation in the metallic compound ${\mathrm{Gd}}_{3}{\mathrm{Ru}}_{4}{\mathrm{Al}}_{12}$ with a distorted kagome lattice structure}},\ }\href@noop {} {\bibfield  {journal} {\bibinfo  {journal} {Phys. Rev. B}\ }\textbf {\bibinfo {volume} {98}},\ \bibinfo {pages} {054410} (\bibinfo {year} {2018})}\BibitemShut {NoStop}%
\bibitem [{\citenamefont {Matsumura}\ \emph {et~al.}(2019)\citenamefont {Matsumura}, \citenamefont {Ozono}, \citenamefont {Nakamura}, \citenamefont {Kabeya},\ and\ \citenamefont {Ochiai}}]{Matsumura2019}%
  \BibitemOpen
  \bibfield  {author} {\bibinfo {author} {\bibfnamefont {T.}~\bibnamefont {Matsumura}}, \bibinfo {author} {\bibfnamefont {Y.}~\bibnamefont {Ozono}}, \bibinfo {author} {\bibfnamefont {S.}~\bibnamefont {Nakamura}}, \bibinfo {author} {\bibfnamefont {N.}~\bibnamefont {Kabeya}},\ and\ \bibinfo {author} {\bibfnamefont {A.}~\bibnamefont {Ochiai}},\ }\bibfield  {title} {\bibinfo {title} {{Helical ordering of spin trimers in a distorted Kagom\'{e} Lattice of Gd$_3$Ru$_4$Al$_{12}$ studied by resonant X-ray diffraction}},\ }\href@noop {} {\bibfield  {journal} {\bibinfo  {journal} {J. Phys. Soc. Jpn.}\ }\textbf {\bibinfo {volume} {88}},\ \bibinfo {pages} {023704} (\bibinfo {year} {2019})}\BibitemShut {NoStop}%
\bibitem [{\citenamefont {Lovesey}\ and\ \citenamefont {Collins}(1996)}]{REXS_textbook}%
  \BibitemOpen
  \bibfield  {author} {\bibinfo {author} {\bibfnamefont {S.~W.}\ \bibnamefont {Lovesey}}\ and\ \bibinfo {author} {\bibfnamefont {S.~P.}\ \bibnamefont {Collins}},\ }\href@noop {} {\emph {\bibinfo {title} {{X-ray scattering and absorption by magnetic materials}}}},\ \bibinfo {series} {Oxford series on synchrotron radiation}\ No.~\bibinfo {number} {1}\ (\bibinfo  {publisher} {Clarendon Press,Oxford University Press},\ \bibinfo {year} {1996})\BibitemShut {NoStop}%
\bibitem [{\citenamefont {McGuire}\ and\ \citenamefont {Potter}(1975)}]{McGuire1975}%
  \BibitemOpen
  \bibfield  {author} {\bibinfo {author} {\bibfnamefont {T.}~\bibnamefont {McGuire}}\ and\ \bibinfo {author} {\bibfnamefont {R.}~\bibnamefont {Potter}},\ }\bibfield  {title} {\bibinfo {title} {{Anisotropic magnetoresistance in ferromagnetic $3d$ alloys}},\ }\href@noop {} {\bibfield  {journal} {\bibinfo  {journal} {IEEE Transactions on Magnetics}\ }\textbf {\bibinfo {volume} {11}},\ \bibinfo {pages} {1018} (\bibinfo {year} {1975})}\BibitemShut {NoStop}%
\bibitem [{\citenamefont {Okumura}\ \emph {et~al.}(2018)\citenamefont {Okumura}, \citenamefont {Kato},\ and\ \citenamefont {Motome}}]{Okumura2018}%
  \BibitemOpen
  \bibfield  {author} {\bibinfo {author} {\bibfnamefont {S.}~\bibnamefont {Okumura}}, \bibinfo {author} {\bibfnamefont {Y.}~\bibnamefont {Kato}},\ and\ \bibinfo {author} {\bibfnamefont {Y.}~\bibnamefont {Motome}},\ }\bibfield  {title} {\bibinfo {title} {{Lock-in of a chiral soliton lattice by itinerant electrons}},\ }\href@noop {} {\bibfield  {journal} {\bibinfo  {journal} {J. Phys. Soc. Jpn.}\ }\textbf {\bibinfo {volume} {87}},\ \bibinfo {pages} {033708} (\bibinfo {year} {2018})}\BibitemShut {NoStop}%
\bibitem [{\citenamefont {Hodt}\ \emph {et~al.}(2024)\citenamefont {Hodt}, \citenamefont {Bentmann},\ and\ \citenamefont {Linder}}]{Hodt2024}%
  \BibitemOpen
  \bibfield  {author} {\bibinfo {author} {\bibfnamefont {E.~W.}\ \bibnamefont {Hodt}}, \bibinfo {author} {\bibfnamefont {H.}~\bibnamefont {Bentmann}},\ and\ \bibinfo {author} {\bibfnamefont {J.}~\bibnamefont {Linder}},\ }\bibfield  {title} {\bibinfo {title} {{The fate of $p$-wave spin polarization in helimagnets with Rashba spin-orbit coupling}},\ }\href@noop {} {\bibfield  {journal} {\bibinfo  {journal} {arXiv preprint arXiv:2412.12246}\ } (\bibinfo {year} {2024})}\BibitemShut {NoStop}%
\bibitem [{\citenamefont {Nagaosa}\ \emph {et~al.}(2010)\citenamefont {Nagaosa}, \citenamefont {Sinova}, \citenamefont {Onoda}, \citenamefont {MacDonald},\ and\ \citenamefont {Ong}}]{Nagaosa2010}%
  \BibitemOpen
  \bibfield  {author} {\bibinfo {author} {\bibfnamefont {N.}~\bibnamefont {Nagaosa}}, \bibinfo {author} {\bibfnamefont {J.}~\bibnamefont {Sinova}}, \bibinfo {author} {\bibfnamefont {S.}~\bibnamefont {Onoda}}, \bibinfo {author} {\bibfnamefont {A.~H.}\ \bibnamefont {MacDonald}},\ and\ \bibinfo {author} {\bibfnamefont {N.~P.}\ \bibnamefont {Ong}},\ }\bibfield  {title} {\bibinfo {title} {{Anomalous Hall effect}},\ }\href@noop {} {\bibfield  {journal} {\bibinfo  {journal} {Rev. Mod. Phys.}\ }\textbf {\bibinfo {volume} {82}},\ \bibinfo {pages} {1539} (\bibinfo {year} {2010})}\BibitemShut {NoStop}%
\bibitem [{\citenamefont {Nakatsuji}\ \emph {et~al.}(2015)\citenamefont {Nakatsuji}, \citenamefont {Kiyohara},\ and\ \citenamefont {Higo}}]{Nakatsuji2015}%
  \BibitemOpen
  \bibfield  {author} {\bibinfo {author} {\bibfnamefont {S.}~\bibnamefont {Nakatsuji}}, \bibinfo {author} {\bibfnamefont {N.}~\bibnamefont {Kiyohara}},\ and\ \bibinfo {author} {\bibfnamefont {T.}~\bibnamefont {Higo}},\ }\bibfield  {title} {\bibinfo {title} {{Large anomalous Hall effect in a non-collinear antiferromagnet at room temperature}},\ }\href@noop {} {\bibfield  {journal} {\bibinfo  {journal} {Nature}\ }\textbf {\bibinfo {volume} {527}},\ \bibinfo {pages} {212} (\bibinfo {year} {2015})}\BibitemShut {NoStop}%
\bibitem [{\citenamefont {Nayak}\ \emph {et~al.}(2016)\citenamefont {Nayak}, \citenamefont {Fischer}, \citenamefont {Sun}, \citenamefont {Yan}, \citenamefont {Karel}, \citenamefont {Komarek}, \citenamefont {Shekhar}, \citenamefont {Kumar}, \citenamefont {Schnelle}, \citenamefont {Kübler}, \citenamefont {Felser},\ and\ \citenamefont {Parkin}}]{Nayak2016}%
  \BibitemOpen
  \bibfield  {author} {\bibinfo {author} {\bibfnamefont {A.~K.}\ \bibnamefont {Nayak}}, \bibinfo {author} {\bibfnamefont {J.~E.}\ \bibnamefont {Fischer}}, \bibinfo {author} {\bibfnamefont {Y.}~\bibnamefont {Sun}}, \bibinfo {author} {\bibfnamefont {B.}~\bibnamefont {Yan}}, \bibinfo {author} {\bibfnamefont {J.}~\bibnamefont {Karel}}, \bibinfo {author} {\bibfnamefont {A.~C.}\ \bibnamefont {Komarek}}, \bibinfo {author} {\bibfnamefont {C.}~\bibnamefont {Shekhar}}, \bibinfo {author} {\bibfnamefont {N.}~\bibnamefont {Kumar}}, \bibinfo {author} {\bibfnamefont {W.}~\bibnamefont {Schnelle}}, \bibinfo {author} {\bibfnamefont {J.}~\bibnamefont {Kübler}}, \bibinfo {author} {\bibfnamefont {C.}~\bibnamefont {Felser}},\ and\ \bibinfo {author} {\bibfnamefont {S.~S.~P.}\ \bibnamefont {Parkin}},\ }\bibfield  {title} {\bibinfo {title} {{Large anomalous Hall effect driven by a nonvanishing Berry curvature in the noncolinear antiferromagnet $\mathrm{Mn}_{3}\mathrm{Ge}$}},\ }\href@noop {} {\bibfield  {journal} {\bibinfo  {journal}
  {Sci. Adv.}\ }\textbf {\bibinfo {volume} {2}},\ \bibinfo {pages} {e1501870} (\bibinfo {year} {2016})}\BibitemShut {NoStop}%
\bibitem [{\citenamefont {Ghimire}\ \emph {et~al.}(2018)\citenamefont {Ghimire}, \citenamefont {Botana}, \citenamefont {Jiang}, \citenamefont {Zhang}, \citenamefont {Chen},\ and\ \citenamefont {Mitchell}}]{Ghimire2018}%
  \BibitemOpen
  \bibfield  {author} {\bibinfo {author} {\bibfnamefont {N.~J.}\ \bibnamefont {Ghimire}}, \bibinfo {author} {\bibfnamefont {A.~S.}\ \bibnamefont {Botana}}, \bibinfo {author} {\bibfnamefont {J.~S.}\ \bibnamefont {Jiang}}, \bibinfo {author} {\bibfnamefont {J.}~\bibnamefont {Zhang}}, \bibinfo {author} {\bibfnamefont {Y.-S.}\ \bibnamefont {Chen}},\ and\ \bibinfo {author} {\bibfnamefont {J.~F.}\ \bibnamefont {Mitchell}},\ }\bibfield  {title} {\bibinfo {title} {{Large anomalous Hall effect in the chiral-lattice antiferromagnet $\mathrm{CoNb}_{3}\mathrm{S}_{6}$}},\ }\href@noop {} {\bibfield  {journal} {\bibinfo  {journal} {Nat. Commun.}\ }\textbf {\bibinfo {volume} {9}},\ \bibinfo {pages} {3280} (\bibinfo {year} {2018})}\BibitemShut {NoStop}%
\bibitem [{\citenamefont {Takagi}\ \emph {et~al.}(2023)\citenamefont {Takagi}, \citenamefont {Takagi}, \citenamefont {Minami}, \citenamefont {Nomoto}, \citenamefont {Ohishi}, \citenamefont {Suzuki}, \citenamefont {Yanagi}, \citenamefont {Hirayama}, \citenamefont {Khanh}, \citenamefont {Karube}, \citenamefont {Saito}, \citenamefont {Hashizume}, \citenamefont {Kiyanagi}, \citenamefont {Tokura}, \citenamefont {Arita}, \citenamefont {Nakajima},\ and\ \citenamefont {Seki}}]{Takagi2023}%
  \BibitemOpen
  \bibfield  {author} {\bibinfo {author} {\bibfnamefont {H.}~\bibnamefont {Takagi}}, \bibinfo {author} {\bibfnamefont {R.}~\bibnamefont {Takagi}}, \bibinfo {author} {\bibfnamefont {S.}~\bibnamefont {Minami}}, \bibinfo {author} {\bibfnamefont {T.}~\bibnamefont {Nomoto}}, \bibinfo {author} {\bibfnamefont {K.}~\bibnamefont {Ohishi}}, \bibinfo {author} {\bibfnamefont {M.-T.}\ \bibnamefont {Suzuki}}, \bibinfo {author} {\bibfnamefont {Y.}~\bibnamefont {Yanagi}}, \bibinfo {author} {\bibfnamefont {M.}~\bibnamefont {Hirayama}}, \bibinfo {author} {\bibfnamefont {N.~D.}\ \bibnamefont {Khanh}}, \bibinfo {author} {\bibfnamefont {K.}~\bibnamefont {Karube}}, \bibinfo {author} {\bibfnamefont {H.}~\bibnamefont {Saito}}, \bibinfo {author} {\bibfnamefont {D.}~\bibnamefont {Hashizume}}, \bibinfo {author} {\bibfnamefont {R.}~\bibnamefont {Kiyanagi}}, \bibinfo {author} {\bibfnamefont {Y.}~\bibnamefont {Tokura}}, \bibinfo {author} {\bibfnamefont {R.}~\bibnamefont {Arita}}, \bibinfo {author} {\bibfnamefont {T.}~\bibnamefont
  {Nakajima}},\ and\ \bibinfo {author} {\bibfnamefont {S.}~\bibnamefont {Seki}},\ }\bibfield  {title} {\bibinfo {title} {{Spontaneous topological Hall effect induced by non-coplanar antiferromagnetic order in intercalated van der Waals materials}},\ }\href@noop {} {\bibfield  {journal} {\bibinfo  {journal} {Nat. Phys.}\ }\textbf {\bibinfo {volume} {19}},\ \bibinfo {pages} {961} (\bibinfo {year} {2023})}\BibitemShut {NoStop}%
\bibitem [{\citenamefont {Park}\ \emph {et~al.}(2023)\citenamefont {Park}, \citenamefont {Cho}, \citenamefont {Kim}, \citenamefont {An}, \citenamefont {Kang}, \citenamefont {Avdeev}, \citenamefont {Sibille}, \citenamefont {Iida}, \citenamefont {Kajimoto}, \citenamefont {Lee}, \citenamefont {Ju}, \citenamefont {Cho}, \citenamefont {Noh}, \citenamefont {Han}, \citenamefont {Zhang}, \citenamefont {Batista},\ and\ \citenamefont {Park}}]{Park2023}%
  \BibitemOpen
  \bibfield  {author} {\bibinfo {author} {\bibfnamefont {P.}~\bibnamefont {Park}}, \bibinfo {author} {\bibfnamefont {W.}~\bibnamefont {Cho}}, \bibinfo {author} {\bibfnamefont {C.}~\bibnamefont {Kim}}, \bibinfo {author} {\bibfnamefont {Y.}~\bibnamefont {An}}, \bibinfo {author} {\bibfnamefont {Y.-G.}\ \bibnamefont {Kang}}, \bibinfo {author} {\bibfnamefont {M.}~\bibnamefont {Avdeev}}, \bibinfo {author} {\bibfnamefont {R.}~\bibnamefont {Sibille}}, \bibinfo {author} {\bibfnamefont {K.}~\bibnamefont {Iida}}, \bibinfo {author} {\bibfnamefont {R.}~\bibnamefont {Kajimoto}}, \bibinfo {author} {\bibfnamefont {K.~H.}\ \bibnamefont {Lee}}, \bibinfo {author} {\bibfnamefont {W.}~\bibnamefont {Ju}}, \bibinfo {author} {\bibfnamefont {E.-J.}\ \bibnamefont {Cho}}, \bibinfo {author} {\bibfnamefont {H.-J.}\ \bibnamefont {Noh}}, \bibinfo {author} {\bibfnamefont {M.~J.}\ \bibnamefont {Han}}, \bibinfo {author} {\bibfnamefont {S.-S.}\ \bibnamefont {Zhang}}, \bibinfo {author} {\bibfnamefont {C.~D.}\ \bibnamefont {Batista}},\ and\
  \bibinfo {author} {\bibfnamefont {J.-G.}\ \bibnamefont {Park}},\ }\bibfield  {title} {\bibinfo {title} {{Tetrahedral triple-Q magnetic ordering and large spontaneous Hall conductivity in the metallic triangular antiferromagnet $\mathrm{Co}_{1/3}\mathrm{TaS}_{2}$}},\ }\href@noop {} {\bibfield  {journal} {\bibinfo  {journal} {Nat. Commun.}\ }\textbf {\bibinfo {volume} {14}},\ \bibinfo {pages} {8346} (\bibinfo {year} {2023})}\BibitemShut {NoStop}%
\bibitem [{\citenamefont {\ifmmode~\check{S}\else \v{S}\fi{}mejkal}\ \emph {et~al.}(2022{\natexlab{c}})\citenamefont {\ifmmode~\check{S}\else \v{S}\fi{}mejkal}, \citenamefont {MacDonald}, \citenamefont {Sinova}, \citenamefont {Nakatsuji},\ and\ \citenamefont {Jungwirth}}]{Smejkal2022NRM}%
  \BibitemOpen
  \bibfield  {author} {\bibinfo {author} {\bibfnamefont {L.}~\bibnamefont {\ifmmode~\check{S}\else \v{S}\fi{}mejkal}}, \bibinfo {author} {\bibfnamefont {A.~H.}\ \bibnamefont {MacDonald}}, \bibinfo {author} {\bibfnamefont {J.}~\bibnamefont {Sinova}}, \bibinfo {author} {\bibfnamefont {S.}~\bibnamefont {Nakatsuji}},\ and\ \bibinfo {author} {\bibfnamefont {T.}~\bibnamefont {Jungwirth}},\ }\bibfield  {title} {\bibinfo {title} {{Anomalous Hall antiferromagnets}},\ }\href@noop {} {\bibfield  {journal} {\bibinfo  {journal} {Nat. Rev. Mater.}\ }\textbf {\bibinfo {volume} {7}},\ \bibinfo {pages} {482} (\bibinfo {year} {2022}{\natexlab{c}})}\BibitemShut {NoStop}%
\bibitem [{\citenamefont {Brekke}\ \emph {et~al.}(2024{\natexlab{b}})\citenamefont {Brekke}, \citenamefont {Sukhachov}, \citenamefont {Giil}, \citenamefont {Brataas},\ and\ \citenamefont {Linder}}]{Brekke2024}%
  \BibitemOpen
  \bibfield  {author} {\bibinfo {author} {\bibfnamefont {B.}~\bibnamefont {Brekke}}, \bibinfo {author} {\bibfnamefont {P.}~\bibnamefont {Sukhachov}}, \bibinfo {author} {\bibfnamefont {H.~G.}\ \bibnamefont {Giil}}, \bibinfo {author} {\bibfnamefont {A.}~\bibnamefont {Brataas}},\ and\ \bibinfo {author} {\bibfnamefont {J.}~\bibnamefont {Linder}},\ }\bibfield  {title} {\bibinfo {title} {{Minimal models and transport properties of unconventional $p$-wave magnets}},\ }\href@noop {} {\bibfield  {journal} {\bibinfo  {journal} {arXiv preprint arXiv:2405.15823}\ } (\bibinfo {year} {2024}{\natexlab{b}})}\BibitemShut {NoStop}%
\bibitem [{\citenamefont {Hedayati}\ and\ \citenamefont {Salehi}(2024)}]{Hedayati2024}%
  \BibitemOpen
  \bibfield  {author} {\bibinfo {author} {\bibfnamefont {A.~A.}\ \bibnamefont {Hedayati}}\ and\ \bibinfo {author} {\bibfnamefont {M.}~\bibnamefont {Salehi}},\ }\bibfield  {title} {\bibinfo {title} {{Transverse Spin current at the normal/p-wave altermagnet junctions}},\ }\href@noop {} {\bibfield  {journal} {\bibinfo  {journal} {arXiv preprint arXiv:2408.10413}\ } (\bibinfo {year} {2024})}\BibitemShut {NoStop}%
\bibitem [{\citenamefont {Álvarez Pari}\ \emph {et~al.}(2024)\citenamefont {Álvarez Pari}, \citenamefont {Jaeschke-Ubiergo}, \citenamefont {Chakraborty}, \citenamefont {Šmejkal},\ and\ \citenamefont {Sinova}}]{Pari2024}%
  \BibitemOpen
  \bibfield  {author} {\bibinfo {author} {\bibfnamefont {N.~A.}\ \bibnamefont {Álvarez Pari}}, \bibinfo {author} {\bibfnamefont {R.}~\bibnamefont {Jaeschke-Ubiergo}}, \bibinfo {author} {\bibfnamefont {A.}~\bibnamefont {Chakraborty}}, \bibinfo {author} {\bibfnamefont {L.}~\bibnamefont {Šmejkal}},\ and\ \bibinfo {author} {\bibfnamefont {J.}~\bibnamefont {Sinova}},\ }\bibfield  {title} {\bibinfo {title} {{Non-relativistic linear Edelstein effect in non-collinear EuIn$_2$As$_2$}},\ }\href@noop {} {\bibfield  {journal} {\bibinfo  {journal} {arXiv preprint arXiv:2412.10984}\ } (\bibinfo {year} {2024})}\BibitemShut {NoStop}%
\bibitem [{\citenamefont {Choy}\ \emph {et~al.}(2011)\citenamefont {Choy}, \citenamefont {Edge}, \citenamefont {Akhmerov},\ and\ \citenamefont {Beenakker}}]{Choy2011}%
  \BibitemOpen
  \bibfield  {author} {\bibinfo {author} {\bibfnamefont {T.~P.}\ \bibnamefont {Choy}}, \bibinfo {author} {\bibfnamefont {J.~M.}\ \bibnamefont {Edge}}, \bibinfo {author} {\bibfnamefont {A.~R.}\ \bibnamefont {Akhmerov}},\ and\ \bibinfo {author} {\bibfnamefont {C.~W.~J.}\ \bibnamefont {Beenakker}},\ }\bibfield  {title} {\bibinfo {title} {{Majorana fermions emerging from magnetic nanoparticles on a superconductor without spin-orbit coupling}},\ }\href@noop {} {\bibfield  {journal} {\bibinfo  {journal} {Phys. Rev. B}\ }\textbf {\bibinfo {volume} {84}},\ \bibinfo {pages} {195442} (\bibinfo {year} {2011})}\BibitemShut {NoStop}%
\bibitem [{\citenamefont {Martin}\ and\ \citenamefont {Morpurgo}(2012)}]{Martin2012}%
  \BibitemOpen
  \bibfield  {author} {\bibinfo {author} {\bibfnamefont {I.}~\bibnamefont {Martin}}\ and\ \bibinfo {author} {\bibfnamefont {A.~F.}\ \bibnamefont {Morpurgo}},\ }\bibfield  {title} {\bibinfo {title} {{Majorana fermions in superconducting helical magnets}},\ }\href@noop {} {\bibfield  {journal} {\bibinfo  {journal} {Phys. Rev. B}\ }\textbf {\bibinfo {volume} {85}},\ \bibinfo {pages} {144505} (\bibinfo {year} {2012})}\BibitemShut {NoStop}%
\bibitem [{\citenamefont {Nadj-Perge}\ \emph {et~al.}(2013)\citenamefont {Nadj-Perge}, \citenamefont {Drozdov}, \citenamefont {Bernevig},\ and\ \citenamefont {Yazdani}}]{NadjPerge2013}%
  \BibitemOpen
  \bibfield  {author} {\bibinfo {author} {\bibfnamefont {S.}~\bibnamefont {Nadj-Perge}}, \bibinfo {author} {\bibfnamefont {I.~K.}\ \bibnamefont {Drozdov}}, \bibinfo {author} {\bibfnamefont {B.~A.}\ \bibnamefont {Bernevig}},\ and\ \bibinfo {author} {\bibfnamefont {A.}~\bibnamefont {Yazdani}},\ }\bibfield  {title} {\bibinfo {title} {{Proposal for realizing Majorana fermions in chains of magnetic atoms on a superconductor}},\ }\href@noop {} {\bibfield  {journal} {\bibinfo  {journal} {Phys. Rev. B}\ }\textbf {\bibinfo {volume} {88}},\ \bibinfo {pages} {020407} (\bibinfo {year} {2013})}\BibitemShut {NoStop}%
\bibitem [{\citenamefont {Klinovaja}\ \emph {et~al.}(2013)\citenamefont {Klinovaja}, \citenamefont {Stano}, \citenamefont {Yazdani},\ and\ \citenamefont {Loss}}]{Klinovaja2013}%
  \BibitemOpen
  \bibfield  {author} {\bibinfo {author} {\bibfnamefont {J.}~\bibnamefont {Klinovaja}}, \bibinfo {author} {\bibfnamefont {P.}~\bibnamefont {Stano}}, \bibinfo {author} {\bibfnamefont {A.}~\bibnamefont {Yazdani}},\ and\ \bibinfo {author} {\bibfnamefont {D.}~\bibnamefont {Loss}},\ }\bibfield  {title} {\bibinfo {title} {{Topological Superconductivity and Majorana Fermions in RKKY Systems}},\ }\href@noop {} {\bibfield  {journal} {\bibinfo  {journal} {Phys. Rev. Lett.}\ }\textbf {\bibinfo {volume} {111}},\ \bibinfo {pages} {186805} (\bibinfo {year} {2013})}\BibitemShut {NoStop}%
\bibitem [{\citenamefont {Maeda}\ \emph {et~al.}(2024)\citenamefont {Maeda}, \citenamefont {Lu}, \citenamefont {Yada},\ and\ \citenamefont {Tanaka}}]{Maeda2024}%
  \BibitemOpen
  \bibfield  {author} {\bibinfo {author} {\bibfnamefont {K.}~\bibnamefont {Maeda}}, \bibinfo {author} {\bibfnamefont {B.}~\bibnamefont {Lu}}, \bibinfo {author} {\bibfnamefont {K.}~\bibnamefont {Yada}},\ and\ \bibinfo {author} {\bibfnamefont {Y.}~\bibnamefont {Tanaka}},\ }\bibfield  {title} {\bibinfo {title} {{Theory of Tunneling Spectroscopy in Unconventional $p$-Wave Magnet-Superconductor Hybrid Structures}},\ }\href@noop {} {\bibfield  {journal} {\bibinfo  {journal} {J. Phys. Soc. Jpn.}\ }\textbf {\bibinfo {volume} {93}},\ \bibinfo {pages} {114703} (\bibinfo {year} {2024})}\BibitemShut {NoStop}%
\bibitem [{\citenamefont {Song}\ \emph {et~al.}(2025)\citenamefont {Song}, \citenamefont {Stavri\'{c}}, \citenamefont {Barone}, \citenamefont {Droghetti}, \citenamefont {Antonenko}, \citenamefont {Venderbos}, \citenamefont {Occhialini}, \citenamefont {Ilyas}, \citenamefont {Erge\c{c}en}, \citenamefont {Gedik}, \citenamefont {Cheong}, \citenamefont {Fernandes}, \citenamefont {Picozzi},\ and\ \citenamefont {Comin}}]{Song2025}%
  \BibitemOpen
  \bibfield  {author} {\bibinfo {author} {\bibfnamefont {Q.}~\bibnamefont {Song}}, \bibinfo {author} {\bibfnamefont {S.}~\bibnamefont {Stavri\'{c}}}, \bibinfo {author} {\bibfnamefont {P.}~\bibnamefont {Barone}}, \bibinfo {author} {\bibfnamefont {A.}~\bibnamefont {Droghetti}}, \bibinfo {author} {\bibfnamefont {D.~S.}\ \bibnamefont {Antonenko}}, \bibinfo {author} {\bibfnamefont {J.~W.~F.}\ \bibnamefont {Venderbos}}, \bibinfo {author} {\bibfnamefont {C.~A.}\ \bibnamefont {Occhialini}}, \bibinfo {author} {\bibfnamefont {B.}~\bibnamefont {Ilyas}}, \bibinfo {author} {\bibfnamefont {E.}~\bibnamefont {Erge\c{c}en}}, \bibinfo {author} {\bibfnamefont {N.}~\bibnamefont {Gedik}}, \bibinfo {author} {\bibfnamefont {S.-W.}\ \bibnamefont {Cheong}}, \bibinfo {author} {\bibfnamefont {R.~M.}\ \bibnamefont {Fernandes}}, \bibinfo {author} {\bibfnamefont {S.}~\bibnamefont {Picozzi}},\ and\ \bibinfo {author} {\bibfnamefont {R.}~\bibnamefont {Comin}},\ }\bibfield  {title} {\bibinfo {title} {{Electrical switching of a $p$-wave
  magnet}},\ }\href@noop {} {\bibfield  {journal} {\bibinfo  {journal} {Nature}\ }\textbf {\bibinfo {volume} {642}},\ \bibinfo {pages} {64} (\bibinfo {year} {2025})}\BibitemShut {NoStop}%
\bibitem [{\citenamefont {Aharoni}(1998)}]{AAharoni1998}%
  \BibitemOpen
  \bibfield  {author} {\bibinfo {author} {\bibfnamefont {A.}~\bibnamefont {Aharoni}},\ }\bibfield  {title} {\bibinfo {title} {{Demagnetizing factors for rectangular ferromagnetic prisms}},\ }\href@noop {} {\bibfield  {journal} {\bibinfo  {journal} {J. Appl. Phys.}\ }\textbf {\bibinfo {volume} {83}},\ \bibinfo {pages} {3432} (\bibinfo {year} {1998})}\BibitemShut {NoStop}%
\bibitem [{\citenamefont {Kresse}\ and\ \citenamefont {Furthm\"uller}(1996)}]{Kresse1996}%
  \BibitemOpen
  \bibfield  {author} {\bibinfo {author} {\bibfnamefont {G.}~\bibnamefont {Kresse}}\ and\ \bibinfo {author} {\bibfnamefont {J.}~\bibnamefont {Furthm\"uller}},\ }\bibfield  {title} {\bibinfo {title} {{Efficient iterative schemes for ab initio total-energy calculations using a plane-wave basis set}},\ }\href@noop {} {\bibfield  {journal} {\bibinfo  {journal} {Phys. Rev. B}\ }\textbf {\bibinfo {volume} {54}},\ \bibinfo {pages} {11169} (\bibinfo {year} {1996})}\BibitemShut {NoStop}%
\bibitem [{\citenamefont {Perdew}\ \emph {et~al.}()\citenamefont {Perdew}, \citenamefont {Burke},\ and\ \citenamefont {Ernzerhof}}]{Perdew1996}%
  \BibitemOpen
  \bibfield  {author} {\bibinfo {author} {\bibfnamefont {J.~P.}\ \bibnamefont {Perdew}}, \bibinfo {author} {\bibfnamefont {K.}~\bibnamefont {Burke}},\ and\ \bibinfo {author} {\bibfnamefont {M.}~\bibnamefont {Ernzerhof}},\ }\bibfield  {title} {\bibinfo {title} {{Generalized Gradient Approximation Made Simple}},\ }\href@noop {} {\bibfield  {journal} {\bibinfo  {journal} {Phys. Rev. Lett.}\ }\textbf {\bibinfo {volume} {77}}}\BibitemShut {NoStop}%
\bibitem [{\citenamefont {Harmon}\ \emph {et~al.}(1995)\citenamefont {Harmon}, \citenamefont {Antropov}, \citenamefont {Liechtenstein}, \citenamefont {Solovyev}, ,\ and\ \citenamefont {Anisimov}}]{BNHarmon}%
  \BibitemOpen
  \bibfield  {author} {\bibinfo {author} {\bibfnamefont {B.}~\bibnamefont {Harmon}}, \bibinfo {author} {\bibfnamefont {V.}~\bibnamefont {Antropov}}, \bibinfo {author} {\bibfnamefont {A.}~\bibnamefont {Liechtenstein}}, \bibinfo {author} {\bibfnamefont {I.}~\bibnamefont {Solovyev}}, ,\ and\ \bibinfo {author} {\bibfnamefont {V.}~\bibnamefont {Anisimov}},\ }\bibfield  {title} {\bibinfo {title} {{Calculation of magneto-optical properties for $4f$ systems: LSDA + Hubbard $U$ results}},\ }\href@noop {} {\bibfield  {journal} {\bibinfo  {journal} {J. Phys. Chem. Solids}\ }\textbf {\bibinfo {volume} {56}} (\bibinfo {year} {1995})}\BibitemShut {NoStop}%
\bibitem [{\citenamefont {Shick}\ \emph {et~al.}(1999)\citenamefont {Shick}, \citenamefont {Liechtenstein},\ and\ \citenamefont {Pickett}}]{ABShick}%
  \BibitemOpen
  \bibfield  {author} {\bibinfo {author} {\bibfnamefont {A.~B.}\ \bibnamefont {Shick}}, \bibinfo {author} {\bibfnamefont {A.~I.}\ \bibnamefont {Liechtenstein}},\ and\ \bibinfo {author} {\bibfnamefont {W.~E.}\ \bibnamefont {Pickett}},\ }\bibfield  {title} {\bibinfo {title} {{Implementation of the LDA+$U$ method using the full-potential linearized augmented plane-wave basis}},\ }\href@noop {} {\bibfield  {journal} {\bibinfo  {journal} {Physical Review B}\ }\textbf {\bibinfo {volume} {60}} (\bibinfo {year} {1999})}\BibitemShut {NoStop}%
\bibitem [{\citenamefont {Ikhlas}\ \emph {et~al.}(2017)\citenamefont {Ikhlas}, \citenamefont {Tomita}, \citenamefont {Koretsune}, \citenamefont {Suzuki}, \citenamefont {Nishio-Hamane}, \citenamefont {Arita}, \citenamefont {Otani},\ and\ \citenamefont {Nakatsuji}}]{Ikhlas2017}%
  \BibitemOpen
  \bibfield  {author} {\bibinfo {author} {\bibfnamefont {M.}~\bibnamefont {Ikhlas}}, \bibinfo {author} {\bibfnamefont {T.}~\bibnamefont {Tomita}}, \bibinfo {author} {\bibfnamefont {T.}~\bibnamefont {Koretsune}}, \bibinfo {author} {\bibfnamefont {M.-T.}\ \bibnamefont {Suzuki}}, \bibinfo {author} {\bibfnamefont {D.}~\bibnamefont {Nishio-Hamane}}, \bibinfo {author} {\bibfnamefont {R.}~\bibnamefont {Arita}}, \bibinfo {author} {\bibfnamefont {Y.}~\bibnamefont {Otani}},\ and\ \bibinfo {author} {\bibfnamefont {S.}~\bibnamefont {Nakatsuji}},\ }\bibfield  {title} {\bibinfo {title} {{Large anomalous Nernst effect at room temperature in a chiral antiferromagnet}},\ }\href@noop {} {\bibfield  {journal} {\bibinfo  {journal} {Nat. Phys.}\ }\textbf {\bibinfo {volume} {13}},\ \bibinfo {pages} {1085} (\bibinfo {year} {2017})}\BibitemShut {NoStop}%
\bibitem [{\citenamefont {Chen}\ \emph {et~al.}(2021)\citenamefont {Chen}, \citenamefont {Tomita}, \citenamefont {Minami}, \citenamefont {Fu}, \citenamefont {Koretsune}, \citenamefont {Kitatani}, \citenamefont {Muhammad}, \citenamefont {Nishio-Hamane}, \citenamefont {Ishii}, \citenamefont {Ishii}, \citenamefont {Arita},\ and\ \citenamefont {Nakatsuji}}]{Chen2021}%
  \BibitemOpen
  \bibfield  {author} {\bibinfo {author} {\bibfnamefont {T.}~\bibnamefont {Chen}}, \bibinfo {author} {\bibfnamefont {T.}~\bibnamefont {Tomita}}, \bibinfo {author} {\bibfnamefont {S.}~\bibnamefont {Minami}}, \bibinfo {author} {\bibfnamefont {M.}~\bibnamefont {Fu}}, \bibinfo {author} {\bibfnamefont {T.}~\bibnamefont {Koretsune}}, \bibinfo {author} {\bibfnamefont {M.}~\bibnamefont {Kitatani}}, \bibinfo {author} {\bibfnamefont {I.}~\bibnamefont {Muhammad}}, \bibinfo {author} {\bibfnamefont {D.}~\bibnamefont {Nishio-Hamane}}, \bibinfo {author} {\bibfnamefont {R.}~\bibnamefont {Ishii}}, \bibinfo {author} {\bibfnamefont {F.}~\bibnamefont {Ishii}}, \bibinfo {author} {\bibfnamefont {R.}~\bibnamefont {Arita}},\ and\ \bibinfo {author} {\bibfnamefont {S.}~\bibnamefont {Nakatsuji}},\ }\bibfield  {title} {\bibinfo {title} {{Anomalous transport due to Weyl fermions in the chiral antiferromagnets $\mathrm{Mn}_{3}X$, $X=\mathrm{Sn}, \mathrm{Ge}$}},\ }\href@noop {} {\bibfield  {journal} {\bibinfo  {journal} {Nat. Commun.}\
  }\textbf {\bibinfo {volume} {12}},\ \bibinfo {pages} {572} (\bibinfo {year} {2021})}\BibitemShut {NoStop}%
\bibitem [{\citenamefont {Liu}\ \emph {et~al.}(2017)\citenamefont {Liu}, \citenamefont {Zhang}, \citenamefont {Liu}, \citenamefont {Ding}, \citenamefont {Liu}, \citenamefont {Jafri}, \citenamefont {Hou}, \citenamefont {Wang}, \citenamefont {Ma},\ and\ \citenamefont {Wu}}]{Liu2017}%
  \BibitemOpen
  \bibfield  {author} {\bibinfo {author} {\bibfnamefont {Z.~H.}\ \bibnamefont {Liu}}, \bibinfo {author} {\bibfnamefont {Y.~J.}\ \bibnamefont {Zhang}}, \bibinfo {author} {\bibfnamefont {G.~D.}\ \bibnamefont {Liu}}, \bibinfo {author} {\bibfnamefont {B.}~\bibnamefont {Ding}}, \bibinfo {author} {\bibfnamefont {E.~K.}\ \bibnamefont {Liu}}, \bibinfo {author} {\bibfnamefont {H.~M.}\ \bibnamefont {Jafri}}, \bibinfo {author} {\bibfnamefont {Z.}~\bibnamefont {Hou}}, \bibinfo {author} {\bibfnamefont {W.~H.}\ \bibnamefont {Wang}}, \bibinfo {author} {\bibfnamefont {X.~Q.}\ \bibnamefont {Ma}},\ and\ \bibinfo {author} {\bibfnamefont {G.~H.}\ \bibnamefont {Wu}},\ }\bibfield  {title} {\bibinfo {title} {{Transition from anomalous Hall effect to topological Hall effect in hexagonal non-collinear magnet Mn$_{3}$Ga}},\ }\href@noop {} {\bibfield  {journal} {\bibinfo  {journal} {{Sci. Rep.}}\ }\textbf {\bibinfo {volume} {7}},\ \bibinfo {pages} {515} (\bibinfo {year} {2017})}\BibitemShut {NoStop}%
\bibitem [{\citenamefont {Hayashi}\ \emph {et~al.}(2023)\citenamefont {Hayashi}, \citenamefont {Shirako}, \citenamefont {Xing}, \citenamefont {Belik}, \citenamefont {Arai}, \citenamefont {Kohno}, \citenamefont {Terashima}, \citenamefont {Kojitani}, \citenamefont {Akaogi},\ and\ \citenamefont {Yamaura}}]{Hayashi2023}%
  \BibitemOpen
  \bibfield  {author} {\bibinfo {author} {\bibfnamefont {H.}~\bibnamefont {Hayashi}}, \bibinfo {author} {\bibfnamefont {Y.}~\bibnamefont {Shirako}}, \bibinfo {author} {\bibfnamefont {L.}~\bibnamefont {Xing}}, \bibinfo {author} {\bibfnamefont {A.~A.}\ \bibnamefont {Belik}}, \bibinfo {author} {\bibfnamefont {M.}~\bibnamefont {Arai}}, \bibinfo {author} {\bibfnamefont {M.}~\bibnamefont {Kohno}}, \bibinfo {author} {\bibfnamefont {T.}~\bibnamefont {Terashima}}, \bibinfo {author} {\bibfnamefont {H.}~\bibnamefont {Kojitani}}, \bibinfo {author} {\bibfnamefont {M.}~\bibnamefont {Akaogi}},\ and\ \bibinfo {author} {\bibfnamefont {K.}~\bibnamefont {Yamaura}},\ }\bibfield  {title} {\bibinfo {title} {{Large anomalous Hall effect observed in the cubic-lattice antiferromagnet ${\mathrm{Mn}}_{3}\text{Sb}$ with kagome lattice}},\ }\href@noop {} {\bibfield  {journal} {\bibinfo  {journal} {Phys. Rev. B}\ }\textbf {\bibinfo {volume} {108}},\ \bibinfo {pages} {075140} (\bibinfo {year} {2023})}\BibitemShut {NoStop}%
\bibitem [{\citenamefont {Zuniga-Cespedes}\ \emph {et~al.}(2023)\citenamefont {Zuniga-Cespedes}, \citenamefont {Manna}, \citenamefont {Noad}, \citenamefont {Yang}, \citenamefont {Nicklas}, \citenamefont {Felser}, \citenamefont {Mackenzie},\ and\ \citenamefont {Hicks}}]{Cespedes2023}%
  \BibitemOpen
  \bibfield  {author} {\bibinfo {author} {\bibfnamefont {B.~E.}\ \bibnamefont {Zuniga-Cespedes}}, \bibinfo {author} {\bibfnamefont {K.}~\bibnamefont {Manna}}, \bibinfo {author} {\bibfnamefont {H.~M.~L.}\ \bibnamefont {Noad}}, \bibinfo {author} {\bibfnamefont {P.-Y.}\ \bibnamefont {Yang}}, \bibinfo {author} {\bibfnamefont {M.}~\bibnamefont {Nicklas}}, \bibinfo {author} {\bibfnamefont {C.}~\bibnamefont {Felser}}, \bibinfo {author} {\bibfnamefont {A.~P.}\ \bibnamefont {Mackenzie}},\ and\ \bibinfo {author} {\bibfnamefont {C.~W.}\ \bibnamefont {Hicks}},\ }\bibfield  {title} {\bibinfo {title} {{Observation of an anomalous Hall effect in single-crystal Mn$_{3}$Pt}},\ }\href@noop {} {\bibfield  {journal} {\bibinfo  {journal} {New J. Phys.}\ }\textbf {\bibinfo {volume} {25}},\ \bibinfo {pages} {023029} (\bibinfo {year} {2023})}\BibitemShut {NoStop}%
\bibitem [{\citenamefont {S{\"u}rgers}\ \emph {et~al.}(2024)\citenamefont {S{\"u}rgers}, \citenamefont {Fischer}, \citenamefont {Campos}, \citenamefont {Hellenes}, \citenamefont {Šmejkal}, \citenamefont {Sinova}, \citenamefont {Merz}, \citenamefont {Wolf},\ and\ \citenamefont {Wernsdorfer}}]{Surgers2024}%
  \BibitemOpen
  \bibfield  {author} {\bibinfo {author} {\bibfnamefont {C.}~\bibnamefont {S{\"u}rgers}}, \bibinfo {author} {\bibfnamefont {G.}~\bibnamefont {Fischer}}, \bibinfo {author} {\bibfnamefont {W.~H.}\ \bibnamefont {Campos}}, \bibinfo {author} {\bibfnamefont {A.~B.}\ \bibnamefont {Hellenes}}, \bibinfo {author} {\bibfnamefont {L.}~\bibnamefont {Šmejkal}}, \bibinfo {author} {\bibfnamefont {J.}~\bibnamefont {Sinova}}, \bibinfo {author} {\bibfnamefont {M.}~\bibnamefont {Merz}}, \bibinfo {author} {\bibfnamefont {T.}~\bibnamefont {Wolf}},\ and\ \bibinfo {author} {\bibfnamefont {W.}~\bibnamefont {Wernsdorfer}},\ }\bibfield  {title} {\bibinfo {title} {{Anomalous Nernst effect in the noncollinear antiferromagnet Mn$_{5}$Si$_{3}$}},\ }\href@noop {} {\bibfield  {journal} {\bibinfo  {journal} {Commun. Mater.}\ }\textbf {\bibinfo {volume} {5}},\ \bibinfo {pages} {176} (\bibinfo {year} {2024})}\BibitemShut {NoStop}%
\bibitem [{\citenamefont {Kotegawa}\ \emph {et~al.}(2023)\citenamefont {Kotegawa}, \citenamefont {Kuwata}, \citenamefont {Huyen}, \citenamefont {Arai}, \citenamefont {Tou}, \citenamefont {Matsuda}, \citenamefont {Takeda}, \citenamefont {Sugawara},\ and\ \citenamefont {Suzuki}}]{Kotegawa2023}%
  \BibitemOpen
  \bibfield  {author} {\bibinfo {author} {\bibfnamefont {H.}~\bibnamefont {Kotegawa}}, \bibinfo {author} {\bibfnamefont {Y.}~\bibnamefont {Kuwata}}, \bibinfo {author} {\bibfnamefont {V.~T.~N.}\ \bibnamefont {Huyen}}, \bibinfo {author} {\bibfnamefont {Y.}~\bibnamefont {Arai}}, \bibinfo {author} {\bibfnamefont {H.}~\bibnamefont {Tou}}, \bibinfo {author} {\bibfnamefont {M.}~\bibnamefont {Matsuda}}, \bibinfo {author} {\bibfnamefont {K.}~\bibnamefont {Takeda}}, \bibinfo {author} {\bibfnamefont {H.}~\bibnamefont {Sugawara}},\ and\ \bibinfo {author} {\bibfnamefont {M.-T.}\ \bibnamefont {Suzuki}},\ }\bibfield  {title} {\bibinfo {title} {{Large anomalous Hall effect and unusual domain switching in an orthorhombic antiferromagnetic material NbMnP}},\ }\href@noop {} {\bibfield  {journal} {\bibinfo  {journal} {npj Quantum Mater.}\ }\textbf {\bibinfo {volume} {8}},\ \bibinfo {pages} {56} (\bibinfo {year} {2023})}\BibitemShut {NoStop}%
\bibitem [{\citenamefont {Kotegawa}\ \emph {et~al.}(2024{\natexlab{a}})\citenamefont {Kotegawa}, \citenamefont {Nakamura}, \citenamefont {Huyen}, \citenamefont {Arai}, \citenamefont {Tou}, \citenamefont {Sugawara}, \citenamefont {Hayashi}, \citenamefont {Takeda}, \citenamefont {Tabata}, \citenamefont {Kaneko}, \citenamefont {Kodama},\ and\ \citenamefont {Suzuki}}]{Kotegawa2024}%
  \BibitemOpen
  \bibfield  {author} {\bibinfo {author} {\bibfnamefont {H.}~\bibnamefont {Kotegawa}}, \bibinfo {author} {\bibfnamefont {A.}~\bibnamefont {Nakamura}}, \bibinfo {author} {\bibfnamefont {V.~T.~N.}\ \bibnamefont {Huyen}}, \bibinfo {author} {\bibfnamefont {Y.}~\bibnamefont {Arai}}, \bibinfo {author} {\bibfnamefont {H.}~\bibnamefont {Tou}}, \bibinfo {author} {\bibfnamefont {H.}~\bibnamefont {Sugawara}}, \bibinfo {author} {\bibfnamefont {J.}~\bibnamefont {Hayashi}}, \bibinfo {author} {\bibfnamefont {K.}~\bibnamefont {Takeda}}, \bibinfo {author} {\bibfnamefont {C.}~\bibnamefont {Tabata}}, \bibinfo {author} {\bibfnamefont {K.}~\bibnamefont {Kaneko}}, \bibinfo {author} {\bibfnamefont {K.}~\bibnamefont {Kodama}},\ and\ \bibinfo {author} {\bibfnamefont {M.-T.}\ \bibnamefont {Suzuki}},\ }\bibfield  {title} {\bibinfo {title} {{Large spontaneous Hall effect with flexible domain control in the antiferromagnetic material TaMnP}},\ }\href@noop {} {\bibfield  {journal} {\bibinfo  {journal} {{Phys. Rev. B}}\ }\textbf {\bibinfo
  {volume} {110}},\ \bibinfo {pages} {214417} (\bibinfo {year} {2024}{\natexlab{a}})}\BibitemShut {NoStop}%
\bibitem [{\citenamefont {Kotegawa}\ \emph {et~al.}(2024{\natexlab{b}})\citenamefont {Kotegawa}, \citenamefont {Tanaka}, \citenamefont {Takeuchi}, \citenamefont {Tou}, \citenamefont {Sugawara}, \citenamefont {Hayashi},\ and\ \citenamefont {Takeda}}]{kotegawa2024PRL}%
  \BibitemOpen
  \bibfield  {author} {\bibinfo {author} {\bibfnamefont {H.}~\bibnamefont {Kotegawa}}, \bibinfo {author} {\bibfnamefont {H.}~\bibnamefont {Tanaka}}, \bibinfo {author} {\bibfnamefont {Y.}~\bibnamefont {Takeuchi}}, \bibinfo {author} {\bibfnamefont {H.}~\bibnamefont {Tou}}, \bibinfo {author} {\bibfnamefont {H.}~\bibnamefont {Sugawara}}, \bibinfo {author} {\bibfnamefont {J.}~\bibnamefont {Hayashi}},\ and\ \bibinfo {author} {\bibfnamefont {K.}~\bibnamefont {Takeda}},\ }\bibfield  {title} {\bibinfo {title} {{Large Anomalous Hall Conductivity Derived from an $f$-Electron Collinear Antiferromagnetic Structure}},\ }\href@noop {} {\bibfield  {journal} {\bibinfo  {journal} {{Phys. Rev. Lett.}}\ }\textbf {\bibinfo {volume} {133}},\ \bibinfo {pages} {106301} (\bibinfo {year} {2024}{\natexlab{b}})}\BibitemShut {NoStop}%
\end{thebibliography}%


\newpage

\newpage
\begin{center}
\Large{Methods}
\end{center}

\textbf{Crystal growth and characterization.} Single crystals of \GRRA{} were synthesized under Ar gas flow in a high-vacuum floating zone (FZ) furnace. The sample quality is confirmed by powder X-ray diffraction (XRD), energy-dispersive X-ray spectroscopy (EDX), and by examination under an optical polarisation microscope. We used Laue XRD and a diamond saw to precisely orient and align the sample surfaces.
\\

\textbf{Magnetization and electrical transport measurements on bulk samples.} Magnetization was measured in a commercial Quantum Design MPMS3 (DC-mode) and Quantum Design PPMS-14T (VSM, vibrating sample magnetometer). The magnetic field is applied along the crystallographic $c$-axis of the crystal. 
Electrical transport measurements were performed on polished single crystal plates with electrical contacts made of $30\,\mathrm{\mu m}$ gold wires, using silver paste (Dupont). We set the sample surface perpendicular to the $c$-axis ($\bm{B} \parallel \bm{c}$) and applied the electric current along the $a$-axis to measure the data in Fig.~\ref{main:Fig3}. We perform a demagnetization correction by assuming that the sample shape is an ellipsis to compare field-dependent physical properties of samples with different shapes~\cite{AAharoni1998}.\\

\textbf{Focused ion beam device fabrication and measurements.} Using a Thermo Fischer Helios 5UX focused ion beam (FIB) system, a lamella was carved and extracted from the bulk crystal, thinned down, and shaped into a circle. Using an in-situ micro-manipulator, this circle was placed on to a Al$_2$O$_3$ substrate patterned with Au contacts, and attached with ion-beam induced Pt deposition. The meander-shaped bonds and final sample geometry were created with a final ion milling step. The entire device was coated in a capping layer of $5\,$nm of Al$_2$O$_3$ using atomic layer deposition. The anisotropic resistance of the finished device was measured in a Quantum Design PPMS-14T as described in the previous Methods paragraph, using the as-provided rotator probe option. \\

\textbf{Resonant elastic X-ray scattering (REXS).} 
Magnetic resonant elastic X-ray scattering (REXS) experiments were conducted at BL-3A (KEK, Photon Factory, Japan) using the Gd-$L_{2}$ absorption edge. The scattering plane is $(H, K, 0)$, so that the incoming and outgoing beam wave vectors $\bm{k}_\mathrm{i}$ and $\bm{k}_\mathrm{f}$ are perpendicular to the crystallographic $c$-axis. We used the $(006)$ reflection of a pyrolytic graphite plate (2$\theta \sim 88^\circ$ at the Gd-$L_{2}$ edge) to perform polarisation analysis of the scattered X-ray beam (see Fig.~\ref{main:Fig2}c and Supplementary Note~\ref{main:SI_REXS} for more details). We define the orientation of the detector with respect to the scattering plane by an angle $\gamma$ such that $\gamma = 0^{\circ}$ detects the $\pi-\sigma'$ component of the scattered X-ray beam, whereas the $\pi-\pi'$ component is detected in the out-of-plane orientation of the detector at $\gamma = 90^{\circ
}$.
\\

\textcolor{black}{\textbf{$p$-wave spin splitting.}}
\textcolor{black}{In this work, we consider collinear spin splitting in the magnetic Brillouin zone of momentum space, i.e., spin-split electronic states bands with a single non-zero component of the spin expectation value $\left<S_x\right> \neq 0$, $\left<S_y\right> = \left<S_z\right> = 0$. $\left<S_x\right> > 0$ and $\left<S_x\right> < 0$ are labeled as $\uparrow$ and $\downarrow$, respectively, although $\left|\left<S_x\right>\right|<\hslash/2$ is allowed. In the low-energy limit, the electronic band structure has an energy gap $\sim \cos(l\theta_{\bm{q}})$ between $\uparrow$, $\downarrow$ bands in momentum space, with $l = 0$, $1$, $2$ for $s$, $p$, $d$ wave spin splitting. The azimuthal angle $\theta_{\bm{q}}$ is the angle of $\bm{q}$ relative to the $x$-axis \cite{Wu2007}; $\bm{q} = \bm{k}-\bm{k}_\mathrm{TRIM}$ is measured from the position of the TRIM around which the low-energy model is expanded (Supplementary Note~\ref{main:SI_Pomeranchuk}).}\\

\textbf{Collinear spin expectation value in momentum space.} 
\textcolor{black}{We use the symmetries of the present material with six-fold expanded magnetic unit cell ($N = 6$) to discuss the spin polarisation of conduction electron states in momentum space.
Here, spin-space group symmetries are denoted as $\left[S\parallel R\right]$, where $S$ acts on spin-space and $R$ applies to real space, while operators without brackets are used to refer to specific (matrix) representations of symmetry operations.}
\textcolor{black}{In the present material, the symmetry $[C_{2\perp}\parallel \bm{t}_{1/2}]$ is equivalent to $[C_{2x}\parallel \bm{t}_{1/2}]$ and all bands and eigenstates can be labelled using its symmetry eigenvalues $\lambda_s$.} We introduce the composite operator $U_s \bm{t}_{1/6}$ with the intention of defining the $\lambda_s$: a sixfold spin-rotation around the $x$-axis, $U_s = \exp(i \tfrac{2\pi}{6} \tfrac{S_x}{\hbar})$, combined with a translation $\bm{t}_{1/6} = \bm{a}_o/6+\bm{b}_o/2$. Here, $S_x$ is the spin-$1/2$ operator component. We use the lattice vectors $\bm{a}_o$, and $\bm{b}_o$ of the magnetic unit cell (orthorhombic), with lattice constants $a_o$, $b_o$ -- as in Fig.~\ref{main:Fig2}d. 

Let a Hamiltonian of conduction electrons be exchange-coupled to the magnetic state discussed in the Main Text. 
Without SOC, this Hamiltonian commutes with \textcolor{black}{
 $(U_s \bm{t}_{1/6})^3 = i \tfrac{2 S_x}{\hbar} \bm{t}_{1/2} \equiv \tilde{S}_x$.
The energy eigenstates $\ket{E, \lambda_s}$ are also eigenstates of  $\tilde{S}_x$ with eigenvalue $\lambda_s$. Using that $\tilde{S}_x$ is unitary and $\tilde{S}_x S_y = -S_y \tilde{S}_x$, the spin expectation value of $S_y$ vanishes, $
\left<S_y\right> = \bra{E, \lambda_s}   \tilde{S}_x^\dagger   \tilde{S}_x S_y \ket{E, \lambda_s} 
= \bra{E, \lambda_s}   \lambda_s^*  (-S_y)  \lambda_s \ket{E, \lambda_s} = - \left<S_y\right> \Rightarrow \left<S_y\right> = 0$, which holds at any momentum $\bm{k}$. By the same argument $\left<S_z\right> = 0$, so only $\left<S_x\right> \neq 0$ is possible.  }
\\

\textbf{Low-energy model from symmetries.}
Starting from the point-group symmetries of $p$-wave magnets, we construct a low-energy model describing an electronic band in two dimensions, coupled to the magnetic texture. \textcolor{black}{The model represents the low-energy limit of the electronic structure of a $p$-wave magnet in the magnetic Brillouin zone.}
\textcolor{black}{For such an expansion around $\bm{k} = 0$, the symmetries $[C_{2x}\parallel \bm{t}_{1/2}]$ and $[\mathcal{T}\parallel \bm{t}_{1/2}]$ are represented by $C_{2x} = i \sigma_x$ and $\mathcal{T} = i \sigma_y K$, respectively, with the complex conjugation $K$. }
Without SOC, the Hamiltonian is required to commute with $C_{2\textcolor{black}{x}}$, and further, is required to be consistent with $p$-wave time-reversal symmetry $\mathcal{T}^{-1} H(\bm{k}) \mathcal{T} = H(-\bm{k})$. To linear order, all allowed terms are equivalent to $p k_x \sigma_x$.
As for spin-orbit coupling (SOC), we consider leading-order terms that are allowed due to broken inversion symmetry \textcolor{black}{and breaking of all mirror planes} in the $p$-wave state. The spin rotation $C_{2x}$ can now be broken and any term of the form $k_i \sigma_j$ with $i \in \{ x, y\}$ and $j \in \{ x, y, z \}$ is consistent with $\mathcal{T}$. However, SOC does not break a combined rotation \textcolor{black}{$[C_{2x}\parallel C_{2x} \bm{t}_{1/2}]$} in spin and direct spaces, which is characteristic for the magnetic state depicted in Fig.~\ref{main:Fig2}i. 
\textcolor{black}{Among the remaining symmetry-allowed terms, $k_x \sigma_y, k_y \sigma_y, k_x \sigma_z, k_y \sigma_z$, the $k_x$-dependent terms are expected to be negligible compared to the $p$-wave splitting. 
As $k_y \sigma_z$ and $k_y \sigma_y$ are equivalent in their effect on spin-splitting, we include only $k_y \sigma_y$, which is needed for a non-zero anomalous Hall effect.}
Finally, the kinetic energy of a parabolic band is $k^2\sigma_0$ and the \textcolor{black}{Zeeman-type energy in an exchange field along the $z$-axis
is $\propto \sigma_z$}. These considerations lead to Eq.~\eqref{main:TwoBandPerturb}\textcolor{black}{, which can be rewritten in terms of the vector $\bm{d}$ as}
\begin{equation}
    \label{main:full_hamiltonian}
    \mathcal{H}(\bm{k}) = t_0\bm{k}^2 \sigma_{0} + p k_{x} \sigma_{x} + m_z \sigma_z + \lambda k_y \sigma_y=t_0\bm{k}^2 \sigma_{0} + \sum_\mathrm{i} d_\mathrm{i} \sigma_\mathrm{i}.
\end{equation}
\textcolor{black}{with eigenenergies $E_\pm = \vb{k}^2 \pm \vert \vb{d} \vert$ and Berry curvature $\Omega_{xy}^{\pm} = \mp \epsilon_{abc}d_{a}\partial_{k_{x}} d_{b} \partial_{k_{y}} d_{c} / (2\left|\bm{d}\right|^3)$, where $\epsilon_{abc}$ is the Levi-Civita symbol.} \textcolor{black}{We note that $\mathcal{H}_0(\bm{k}) = t_0\bm{k}^2 \sigma_{0} + p k_{x} \sigma_{x}$ can also be obtained explicitly as the low-energy limit of a tight-binding model considering conduction electrons coupled to a commensurate spin helix composed of $4f$ local magnetic moments. This calculation is shown in Supplementary Notes~\ref{main:SI_LowEderive}, \ref{main:SI_LowE_perturbative}.}\\ 

\textbf{Spin-nodal plane in momentum space.} In the following, we make some more general remarks on the electronic structure of the $p$-wave magnet. We can argue for the existence of degenerate nodal planes in momentum space by invoking \textcolor{black}{$\mathcal{T}$}
symmetry on the eigenstates of $\tilde{S}_x = \left(U_s \bm{t}_{1/6}\right)^3$ with  eigenvalues $\pm \lambda_s$. The $\mathcal{T}$ operation comprises complex conjugation of wavefunctions and a spin flip, so we have $\mathcal{T}\tilde{S}_x=\tilde{S}_x \mathcal{T}$. This yields $E(\bm{k}) = E(-\bm{k})$, since two eigenstates of the Hamiltonian are related by $\mathcal{T}\ket{E, \bm{k}, \lambda_s} \propto \ket{E,-\bm{k}, -\lambda_s}$ \textcolor{black}{with $\mathcal{T} \lambda_s = -  \lambda_s \mathcal{T}$}.
The $\lambda_s$ is a good quantum number for all crystal momenta $\bm{k}$; hence, along any line connecting the momenta $\bm{k}$ and $-\bm{k}$, there must be an odd number of exchanges between eigenstates labelled by $+\lambda_s$ and $-\lambda_s$. This implies an odd number of nodal planes on which the energy bands of the Hamiltonian are two-fold degenerate in absence of SOC. \\

\textbf{Pinning of nodal plane to high symmetry directions.} 
We discuss symmetry constraints that force the nodal planes to be flat and spanned by high-symmetry directions in $\bm{k}$-space.
Again invoking the \textcolor{black}{$[C_{2x}\parallel C_{2x} \bm{t}_{1/2}]$} symmetry and combining with \textcolor{black}{$[\mathcal{T}\parallel \bm{t}_{1/2}]$}, we define their product \textcolor{black}{$\mathcal{G} = [C_{2x} \mathcal{T} \parallel C_{2x}]$}. This $\mathcal{G}$ commutes with the Hamiltonian and with \textcolor{black}{$[C_{2x}\parallel \bm{t}_{1/2}]$} up to a full lattice translation. In addition, $\mathcal{G}$ acts on crystal momenta like a mirror operation, $(k_x, k_y, k_z) \rightarrow (-k_x, k_y, k_z)$ \textcolor{black}{and exchanges the sign of $\lambda_s$ by complex conjugation}. Thus, $\mathcal{G}$ enforces a nodal plane at $k_x = 0$, where the eigenvalues $\pm\lambda_s$ label states of degenerate energy. \textcolor{black}{Extended Data Fig.~\ref{main:AMR_domains} and Supplementary Fig.~\ref{main:SFig_device_rotation} discuss the two-fold symmetric anisotropic magnetoresistance attributed to electronic anisotropy and the pinned spin-nodal plane in our $p$-wave magnet.}
\\

\color{black}
\textbf{Electronic structure calculations.} 
We performed spin density functional theory (DFT) calculations for \GRA{}. The calculations were carried out using the projector augmented-wave method as implemented in the Vienna ab initio simulation package~\cite{Kresse1996}, together with the generalized gradient approximation for the exchange-correlation functional proposed by Perdew, Burke, and Ernzerhof~\cite{Perdew1996}. The crystal structure data were adopted from Ref.~\cite{Gladyshevskii1993}. In the electronic structure calculations, we employed a plane-wave energy cutoff of $400 \, \mathrm{eV}$ and a $9 \times 9 \times 9$ Monkhorst-Pack $k$-point grid. All calculations were performed assuming a ferromagnetically ordered state. To account for the strong correlation effects in the Gd 4$f$ orbitals, we used the DFT+$U$ approach, setting the on-site Coulomb interaction $U$ to $6.7 \, \mathrm{eV}$ and the Hund's coupling $J$ to $0.7 \, \mathrm{eV}$\cite{BNHarmon, ABShick}.\\
\color{black}

\textbf{Acknowledgments}\\
We acknowledge Akiko Kikkawa for help in crystal growth. We also acknowledge support from the Japan Society for the Promotion of Science (JSPS) under Grant Nos. JP22H04463, JP23H05431, JP22F22742, JP22K20348, JP23K13057, JP24H01607, JP23H00171, 25K17336, and JP24H01604, as well as from the Murata Science Foundation, Yamada Science Foundation, Hattori Hokokai Foundation, Mazda Foundation, Casio Science Promotion Foundation, Inamori Foundation, and Kenjiro Takayanagi Foundation. This work was partially supported by the Japan Science and Technology Agency via JST CREST Grant Numbers JPMJCR1874, JPMJCR20T1, and JPMJCR20T2 (Japan) and JST FOREST (JPMJFR2238). It was also supported by Japan Science and Technology Agency (JST) as part of Adopting Sustainable Partnerships for Innovative Research Ecosystem (ASPIRE), Grant Number JPMJAP2426. P.R.B. acknowledges Swiss National Science Foundation (SNSF) Postdoc.Mobility grant P500PT\_217697 for financial assistance. Mo.~H.~is funded by the Deutsche Forschungsgemeinschaft (DFG, German Research Foundation) - project number 518238332. J.~M.~acknowledges funding from the DFG under the Project No. 547968854. Resonant X-ray scattering at Photon Factory (KEK) was carried out under proposal numbers 2022G551 and 2023G611. The neutron experiments at the Materials and Life Science Experimental Facility of J-PARC were performed under a user program (Proposal No. 2020B0347, TAIKAN, and No. 2020A0282, SENJU).\\

\textbf{Data and code availability}\\
The raw data and code used to produce the figures and conclusions presented in this study have been deposited, with detailed comments, on the \textit{Publication Data Repository System} of RIKEN Center for Emergent Matter Science (Wako, Japan). These data are available from the corresponding authors upon reasonable request. \\

\textbf{Author contributions}\\
Ma.H., Y.Mo., T.A., and Y.To. conceived the project. R.Y. and Ma.H. grew and characterized the single crystals. R.Y., R.N., and Ma.H. performed magnetization and transport measurements on bulk samples. M.T.B. prepared focused ion beam devices and performed all device measurements. P.R.B, R.N., S.G., H.S., H.N., and Ma.H. performed resonant X-ray scattering experiments and analyzed data with guidance from T.-h.A.. T.Na., K.O., K.K., T.O., R.K. and Ma.H. performed neutron scattering experiments, while the data was analyzed by P.R.B. and Y.I. 
S.O., Mo.H., M.E., J.M.,
and Y.Mo. performed symmetry analysis and model calculations. T.No. performed ab-initio calculations. R.Y., M.T.B., P.R.B., and Ma.H. wrote the manuscript with help from I.B.; all authors discussed the results and commented on the manuscript.\\

\textbf{Competing interests}\\
The authors declare no competing interests.


\clearpage

\begin{center}
\Large{Main Text Figures}
\end{center}
\vspace{5mm}
\begin{figure}[htb]%
\centering
\includegraphics[width=0.55\textwidth]{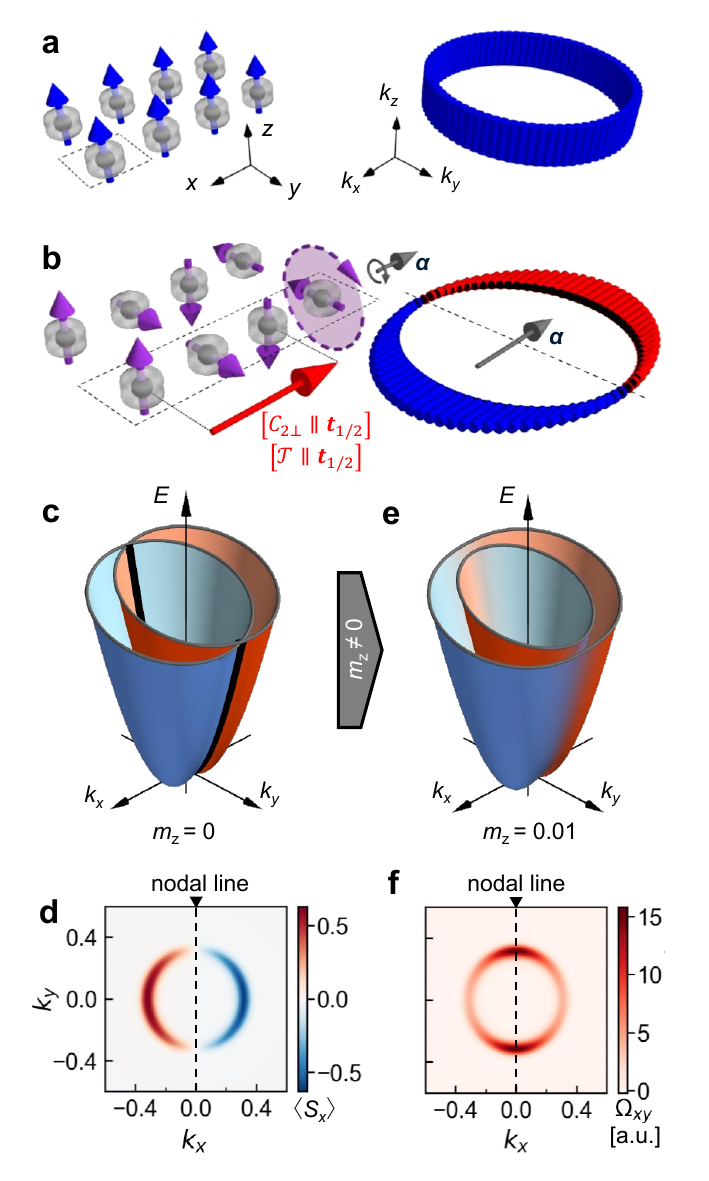}
\caption{\textbf{Magnetic structure of a $p$-wave magnet in direct space and spin-split electronic bands in momentum space.}
\textbf{a}, Ferromagnets have spins along the $z$-axis in direct space and a uniform spin polarisation $S_z(\bm{k})$ in $\bm{k}$-space (right). \textbf{b}, \textcolor{black}{$p$-wave magnetism (conceptual):} Non-collinear spin modulation in direct space and alternating spin polarisation in momentum ($\bm{k}$-)space. Red and blue bars: momentum-resolved expectation value of $S_x(\bm{k})$ at the chemical potential. For commensurate order with wavelength $\lambda_\mathrm{mag} = N$ (even integer $N$), this state is unchanged under $180^\circ$ spin rotation followed by translation of half a lattice constant, termed $[C_{2\perp}\parallel\bm{t}_{1/2}]$ symmetry. The spin polarisation vector is $\bm{\alpha}$~\cite{Ezawa2024_electric_neel}. Violet disk and circular arrow: spin rotation plane in \textcolor{black}{direct} space. \textbf{c}, Minimal electronic structure model for a $p$-wave magnet \textcolor{black}{depicted in the magnetic Brillouin zone according to Eq.~(\ref{main:TwoBandPerturb})}. The spin-splitting of bands causes two-fold electronic anisotropy. Red (blue) indicate spin parallel (antiparallel) to $k_x$. 
\textcolor{black}{\textbf{d}, Spin polarisation $\left<S_x\right>$ when $m_z=\lambda= 0$ in Eq.~\eqref{main:TwoBandPerturb}, which represents a sum of all occupied states -- i.e., of the two bands in (\textbf{c}).}
\textcolor{black}{\textbf{e}, Electronic structure in presence of a tiny net magnetization $m_z$.}
\textbf{f}, Anomalous Hall conductivity $\sigma_{xy}$ induced by Berry curvature $\Omega_{xy}$ appears in the $k_x$-$k_y$ plane when $m_z \neq 0$ \textcolor{black}{and $\lambda \neq 0$}, due to lifting the band degeneracy of the nodal plane. 
}
\label{main:Fig1}
\end{figure}

\clearpage
\begin{figure}[h]%
\centering
\includegraphics[width=1.0\textwidth]{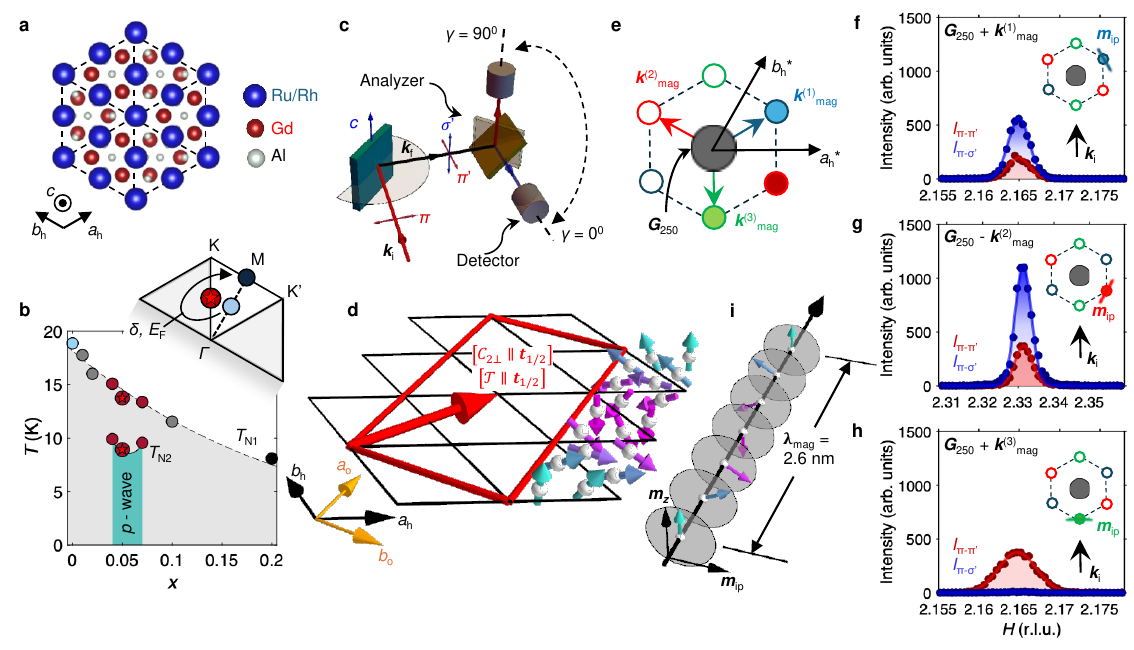}
\caption{\textbf{Commensurate spin helix in direct space, consistent with $p$-wave magnetism, revealed by resonant elastic X-ray scattering (REXS).} 
\textbf{a}, Crystal structure of hexagonal \GRRA{}. 
\textbf{b}, Tuning the magnetic propagation vector $\bm{k}_\mathrm{mag}$ by chemical substitution of Rh for Ru. $T_{\mathrm{N}1}$ (dotted line) is the magnetic transition temperature from the paramagnetic state, while $T_{\mathrm{N}2}$ (solid line) corresponds to the transition into the $p$-wave state.
Inset: Evolution of the magnetic ordering vector $\bm{k}_\mathrm{mag}$ with a shift of the Fermi energy $E_\mathrm{F}$ due to Rh-doping $\delta$: $\bm{k}_\mathrm{mag} = (1/6, 1/6, 0)$ appears for $\delta = 0.04-0.07$, c.f. shaded region in the main panel. For $\delta=0$ and $\delta=0.2$ the magnetic wavevectors in the ground state are $(0.272,0,0)$ and $(0.5,0,0)$, respectively. Momenta are expressed in reciprocal lattice units (r.l.u.) and only half of the first hexagonal Brillouin zone is depicted, sliced at $k_z = 0$.
\textbf{c}, Single crystal REXS experiments in reflection geometry, with incoming and outgoing beams 
$\bm{k}_\mathrm{i}$, $\bm{k}_\mathrm{f}$ spanning the scattering plane (light gray). In polarisation analysis, $I_{\pi- \pi^\prime}$ and $I_{\pi- \sigma^\prime}$ components of $\bm{k}_\mathrm{f}$ are separated by rotating the detector around an analyzer plate by the angle $\gamma$ (Methods).
\textbf{d}, Crystallographic unit cells (hexagonal, black grid) and reconstructed magnetic unit cell (orthorhombic, black box). The requirements for $p$-wave magnetism, $[C_{2\perp}\parallel \bm{t}_{1/2}]$ and $[\mathcal{T}\parallel\bm{t}_{1/2}]$ symmetry, are satisfied by the black arrow.
\textbf{e}, Due to the formation of three $p$-wave magnetic domains, six magnetic reflections appear in a given Brillouin zone. 
\textbf{f}-\textbf{h}, Polarisation analysis using magnetic satellites around the fundamental $(2, 5, 0)$ reflection. Insets: Geometry of $\bm{k}_\mathrm{i}$ relative to the modulated magnetization projected into the scattering plane, $\bm{m}_{\mathrm{ip}}$. 
\textbf{i}, Illustration: Commensurate spin helix with short wavelength $\lambda_\mathrm{mag}$ in direct space. 
}
\label{main:Fig2}
\end{figure}



\clearpage
\begin{figure}[h]%
\centering
\includegraphics[width=1.0\textwidth]{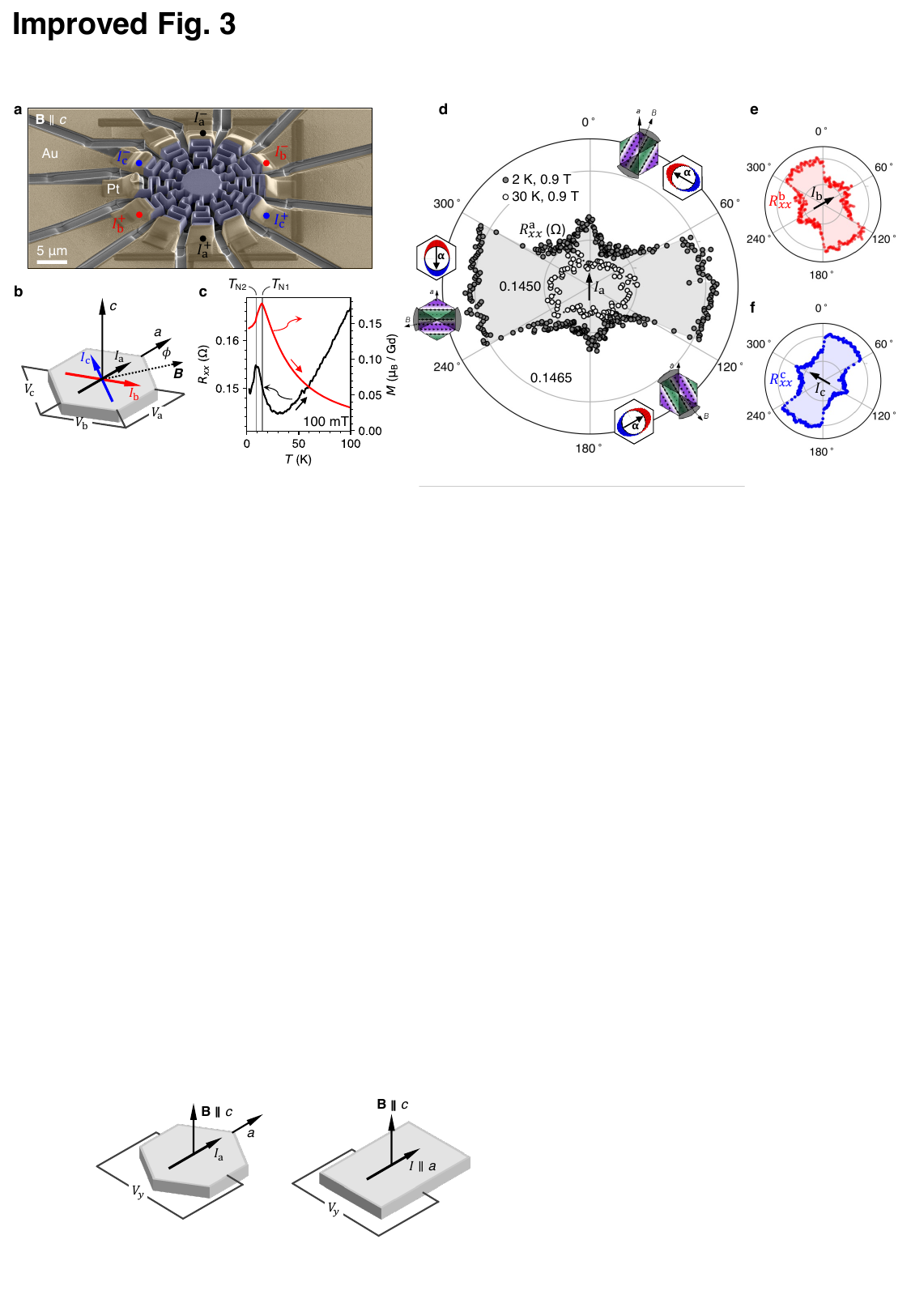}
\caption{\textbf{Anisotropic electronic transport properties.}
\textbf{a}, Scanning electron micrograph of the rotationally symmetric focused ion beam device. The meander-shaped arms release the strain on the device, which arises due to thermal expansion mismatch of device and substrate. The false colour regions indicate the target material (blue) and the electrical contacts (orange), which are composed of evaporated gold (Au) wires and ion beam-deposited platinum (Pt), respectively. The contacts utilised in the directional transport measurements are labelled and colour coded. The scale bar is $5\,$\textmu m. \textbf{b}, Device geometry with crystal axes $a$, $c$ and three pairs of current contacts $I_a$, $I_b$, $I_c$; the corresponding longitudinal voltage drops $V_a$, $V_b$, $V_c$ are measured simultaneously. The magnetic field $\bm{B}$ is rotated by an angle $\phi$ to the $a$-axis. \textbf{c}, Temperature-dependent traces of longitudinal resistance (left) and magnetization $M$ (right), with two antiferromagnetic transitions $T_{\mathrm{N}1}$ and $T_{\mathrm{N}2}$. \textbf{d}, Change in longitudinal resistance $R_{xx}^a$ when switching between different $p$-wave domains by a small magnetic field of $0.9\,$T. Insets: deduced direction of the spin polarisation vector $\bm{\alpha}$ in momentum space, and the periodic spin texture in direct space. The gray wedge indicates the sextant of the magnetic field angle over which the depicted $p$-wave domain is stable. Two-fold anisotropy emerges, with higher resistance for current $I_a$ (center arrow) parallel to $\bm{\alpha}$. Open circles: Data at $30\,\mathrm{K}>T_{\mathrm{N}1}$ plotted for comparison. \textbf{e},\textbf{f} Simultaneous measurements for other contact pairs, demonstrating the anisotropic resistance is due to the relative alignment between current and $\bm{\alpha}$.
}
\label{main:Fig3}
\end{figure}

\clearpage
\begin{figure}[h]%
\centering
\includegraphics[width=1.0\textwidth]{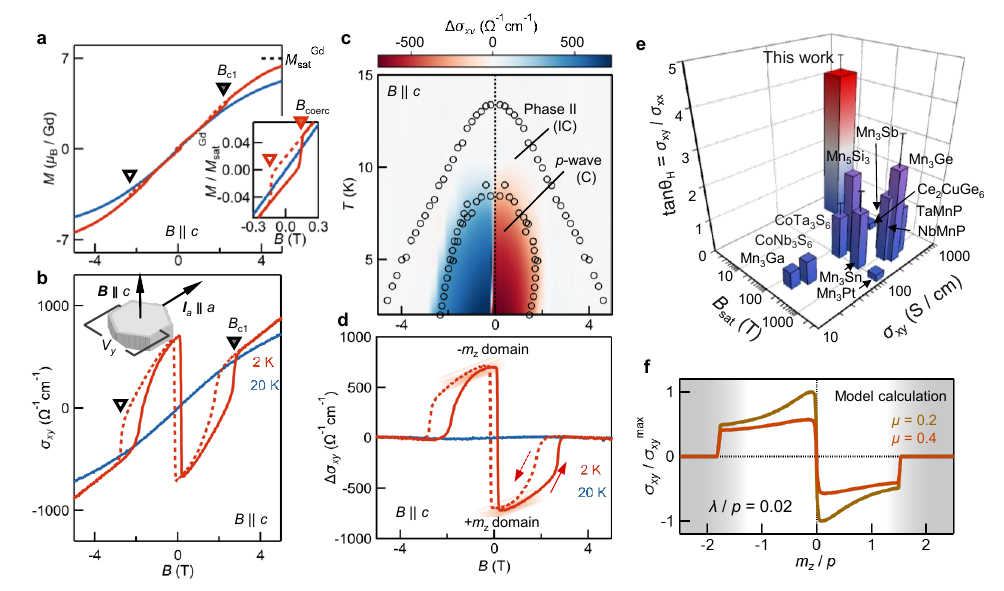}
\caption{\textbf{Giant anomalous Hall effect (AHE) induced by a tiny net magnetization $m_z$ and spin-orbit coupling (SOC).}
\textbf{a}, Comparison of net magnetization $M(B)$ measured at $T= 2\,$K and $20\,$K. The magnetic field is applied along the $c$-axis. Black triangles correspond to a transition field ($B_{\mathrm{c}1}$) from the $p$-wave magnetic phase (commensurate, C) to the high-temperature Phase-II (incommensurate, IC). Inset: small spontaneous magnetization $m_z$ appears around zero field below $B_\mathrm{coerc}$ (red triangles) at low temperature. 
\textbf{b}, Hall conductivity with large zero-field (anomalous) value $\sigma_{xy}^\mathrm{A}$ reaching up to $600 \, \mathrm{\Omega}^{-1}\mathrm{cm}^{-1}$, measured on a bulk single crystal. 
\textbf{c},\textbf{d}, Hall conductivity $\Delta \sigma_{xy}$ after subtraction of components proportional to the magnetic field and to the magnetization (Supplementary Fig.~\ref{main:Efig_AHC_subtraction}). $\Delta \sigma_{xy}$ is enhanced in the $p$-wave magnetic Phase-I. Open circles marking the phase boundaries in panel \textbf{c} are obtained from magnetization measurements in Extended Data Fig.~\ref{main:Efig_Magnetization}. Red arrows in panel \textbf{d} indicate the direction of the field sweep. 
\textbf{e}, As compared to other bulk antiferromagnets with AHE, our slightly distorted $p$-wave state in \GRA{} has large $\sigma_{xy}^\mathrm{A}$ induced by $m_z$, large Hall angle $\sigma_{xy}^\mathrm{A}/ \sigma_{xx}$, and a low saturation magnetic field $B_\mathrm{sat}$. Raw data for the bar plot are provided in Extended Table \ref{main:Table_AHC}.
\textcolor{black}{\textbf{f}, Calculated anomalous Hall conductivity from the low-energy model Eq.~\eqref{main:TwoBandPerturb} plotted against $m_z$/$p$. We consider the breaking of $p$-wave state around $m_z / p \sim 1.5$, where anomalous Hall conductivity is suppressed to zero (see Supplementary Note~\ref{main:conversion_exchange_splitting} for details).}
}
\label{main:Fig4}
\end{figure}

\clearpage
\begin{center}
\Large{Extended Data}
\end{center}

\stepcounter{hoge}
\renewcommand\thefigure{E\arabic{figure}} 
\renewcommand\thetable{E\arabic{table}} 
\renewcommand\thesection{E\arabic{section}} 
\renewcommand\theequation{E\arabic{equation}} 

\makeatletter
\renewcommand\@bibitem[1]{\item\if@filesw \immediate\write\@auxout
    {\string\bibcite{#1}{A\the\value{\@listctr}}}\fi\ignorespaces}
\def\@biblabel#1{E[#1]}
\makeatother

\begin{figure}[h]%
\centering
\includegraphics[width=0.6\textwidth]{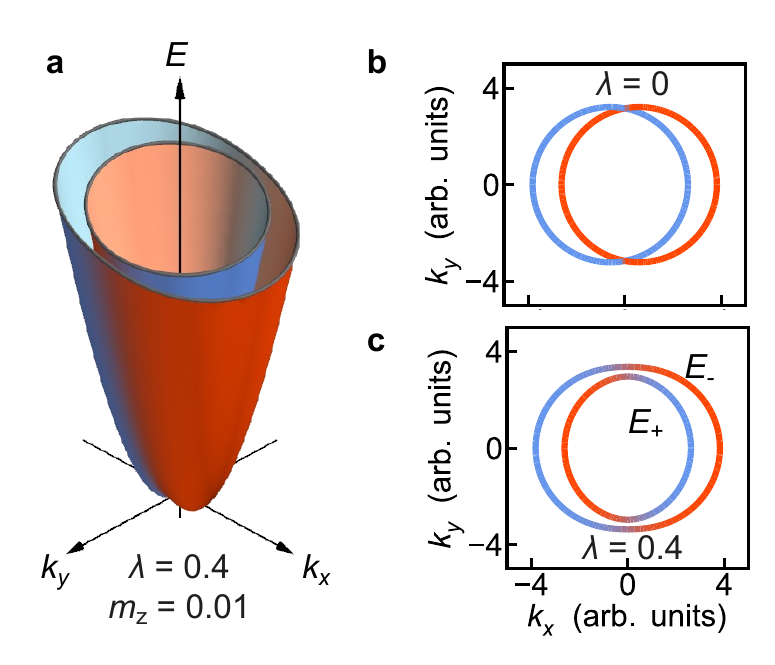}
\caption{\textbf{Fermi surfaces of a $p$-wave band.} 
\textbf{a}, Band dispersion with $p$-wave splitting, spin-orbit coupling (SOC), and time-reversal breaking $m_z = 0.01$. Here, we set $p = 1.2$ and $\lambda = 0.4$, where $p$-wave exchange splitting is larger than the size of SOC and the net magnetization is small but non-zero. \textbf{b},\textbf{c}, Fermi surfaces of a $p$-wave split band with ($m_z = 0$, $\lambda = 0$) and without ($m_z = 0.01$, $\lambda = 0.4$) time-reversal symmetry, respectively. The two bands are degenerate on the $k_x=0$ plane for $m_z = 0$. However, the degeneracy is lifted due to band hybridization when $m_z$
is finite in Eq.~(\ref{main:TwoBandPerturb}). The equal energy surface is calculated at $E = 10$ in panels (\textbf{b}, \textbf{c}). \textcolor{black}{Contrary to the low-energy model for a $p$-wave Pomeranchuk instability, $\left|\left<S_x\right>\right|<\hbar/2$ for the $p$-wave magnet of conduction electrons coupled to a spin helix (Methods, Supplementary Note~\ref{main:SI_Pomeranchuk}).}
}
\label{main:FermiSurface}
\end{figure}

\clearpage
\begin{figure}[h]%
\centering
\includegraphics[width=1.0\textwidth]{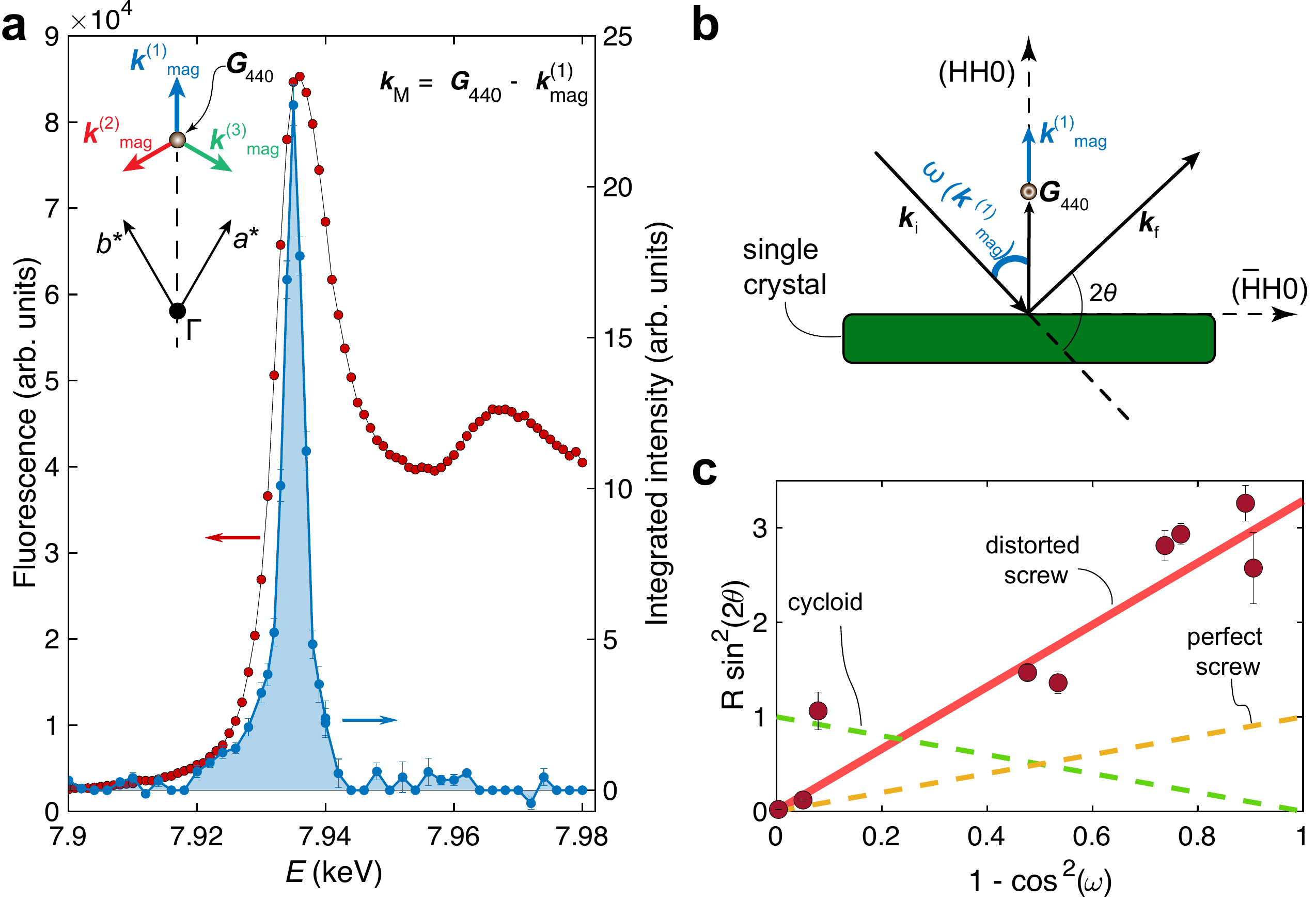}
\caption{\textbf{Resonant elastic X-ray scattering of magnetic satellites in Phase-I}. 
\textbf{a}, The magnetic reflection intensity at $\bm{k}_{\mathrm{M}} (\equiv \bm{G}_{440} - \bm{k}^{(1)}_{\mathrm{mag}}$) and total fluorescence signal obtained at $T = 4\,$K strongly depend on the energy $E$ of the incoming synchrotron X-ray beam. The maximum intensity due to magnetic scattering appears at $E=7.935 \, \mathrm{keV}$, in resonance with the Gd-$L_2$ absorption edge. Inset: schematic of momentum space and magnetic satellites around the $(440)$ fundamental (structural) reflection. We indicate the center of momentum space, $\Gamma$, and lattice vectors of momentum space. \textbf{b}, Measurement geometry for detecting the $\bm{k}_{\mathrm{M}}$ reflection: $\bm{k}_\mathrm{i}$ and $\bm{k}_\mathrm{f}$ are the wavevectors of the incoming and outgoing X-ray beams, respectively. The angles $\omega$, $2\theta$ defined here are referred to in Eq.~(\ref{main:Eq_Rsin2theta_1}) and following. \textbf{c}, polarisation analysis of magnetic satellites close to (440), (250), and (700) fundamental reflections using Eq.~(\ref{main:Eq_Rsin2theta_2}) reveals good agreement with a distorted spin helix model. Dashed green (yellow) lines represent expected behaviour for a harmonic cycloid (harmonic helix). The data are consistent with a distorted helix, elliptically squeezed into the hexagonal basal plane.
}
\label{main:REXS_additional}
\end{figure}

\clearpage
\begin{figure}[h]%
\centering
\includegraphics[width=1.0\textwidth]{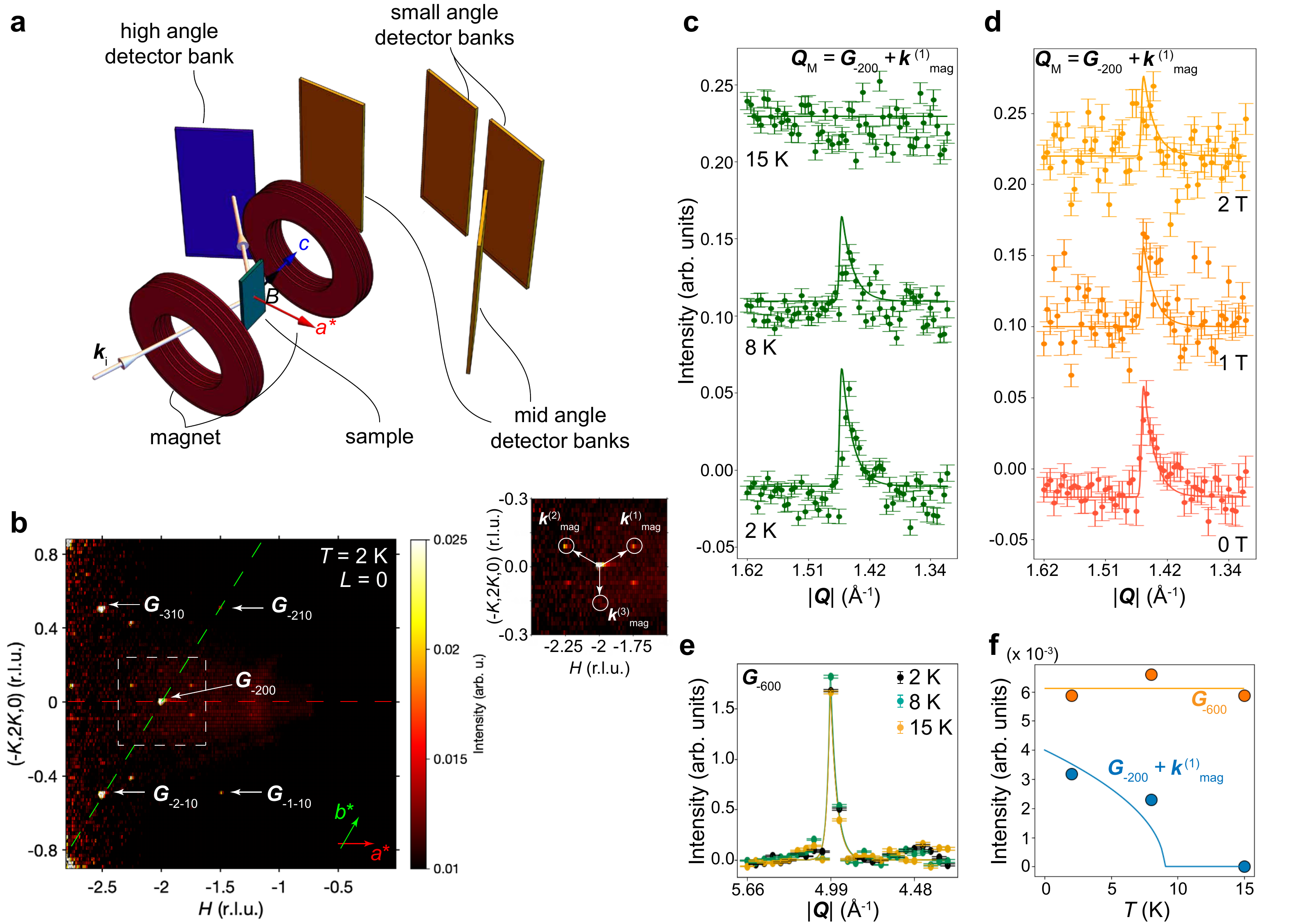}
\caption{\textbf{Magnetic order in Phase-I}, probed by elastic neutron scattering. \textbf{a}, Experimental geometry. A rectangular cuboid shaped crystal with Gd in natural abundance was installed inside a $4\,$T superconducting magnet. The magnet with sample assembly was rotated around the vertical axis by 29$^{\circ}$ in order to access reflections on the high-angle detector. The beamline is equipped with five detector banks; only the high-angle detector (shown in blue) was used to collect scattered neutrons. \textbf{b}, Reciprocal space map at $L = 0$, obtained at $T = 2\,$K and $B = 0\,$T. Data measured on an empty can at $15 \,$K are used for background correction. {\color{black}Nuclear reflections are labelled. The red and green dashed lines represent the coordinate axes in the hexagonal system. Inset: six-fold symmetric magnetic satellite reflections in the region around $\bm{G}_{-200}$, marked by a white dashed box in the main panel. $\bm{k}_{\mathrm{mag}}$-vectors are labelled according to Main Text Fig.~\ref{main:Fig2}e.}
 ({\color{black}C}aption continues on next page.)
}
\label{main:taikan}
\end{figure}

\addtocounter{figure}{-1}
\clearpage
\begin{figure}[h]%
\centering
\caption{{\color{black}(Caption continues from previous page.) {\color{black}\textbf{c},\textbf{d}}, Temperature and magnetic field evolution of $\bm{Q}_\mathrm{M}(\equiv \bm{G}_{-200} + \bm{k}^{(1)}_\mathrm{mag})$ are shown in panels {\color{black}\textbf{c} and \textbf{d}}, respectively. Solid lines represent back-to-back exponential fits as defined in Eq.~(\ref{main:b2b}). The curves are shifted by a constant value for better visibility. 
{\color{black}\textbf{e}}, Line scan through the $\bm{G}_{-600}$ nuclear reflection as a function of total momentum transfer $\left|\bm{Q}\right|$, tracked across the two $T$-induced transitions. \textbf{f}}, Integrated intensities of $\bm{G}_{-600}$ and $\bm{Q}_{\mathrm{M}}$ obtained from fitting a back-to-back exponential function. For the fits shown in panels {\color{black}\textbf{c}, \textbf{d}, and \textbf{e}}, width and center position were kept constant while background and amplitude of the peaks were treated as free parameters. Experiments were performed at beamline BL15 (TAIKAN) of MLF, J-PARC (Tokai, Japan).
}
\end{figure} 

\clearpage

\begin{figure}[h]%
\centering
\includegraphics[width=0.85\textwidth]{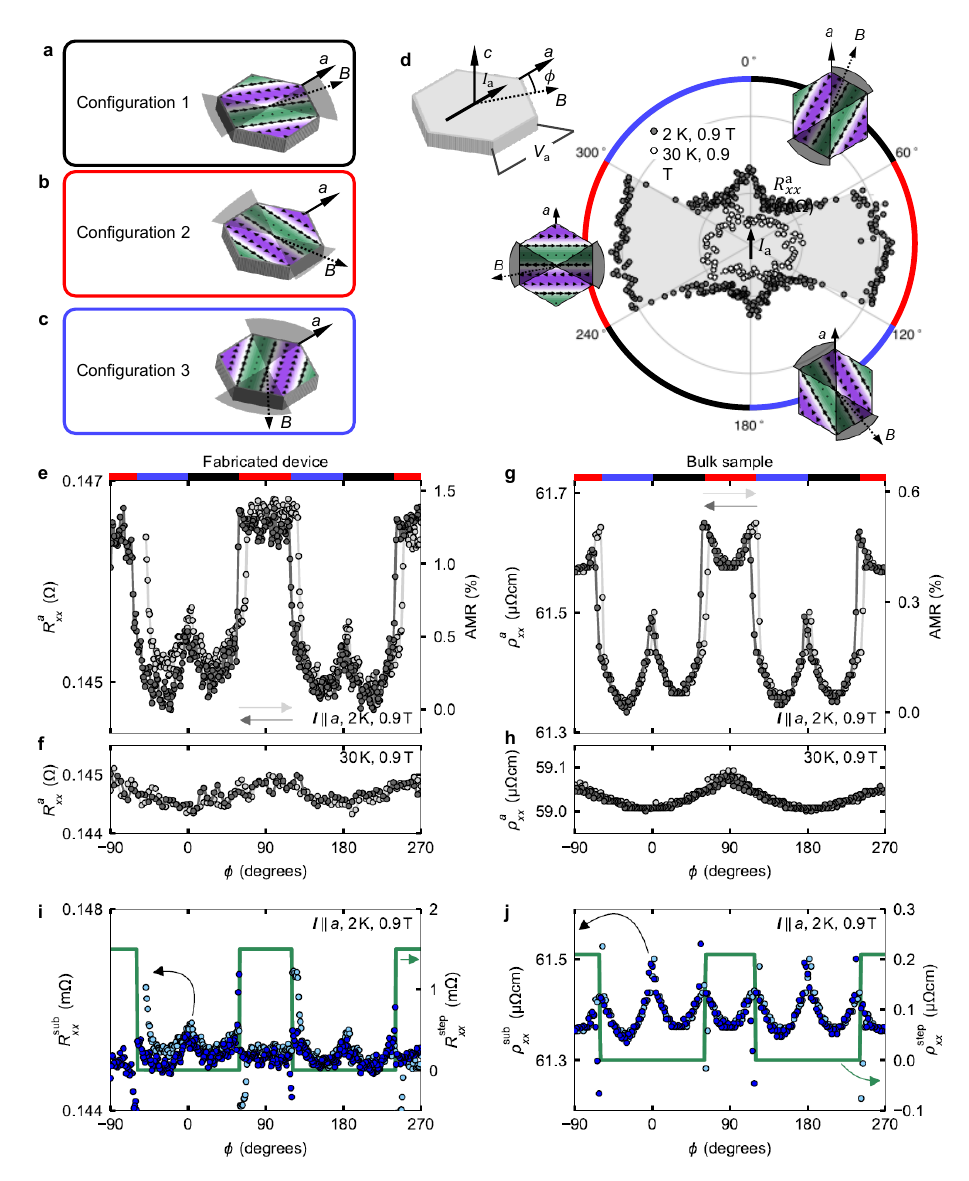}
\caption{\color{black}\textbf{Step-like anisotropic magnetoresistance due to switching of $p$-wave domains, corresponding to switching the spin polarisation vector $\bm{\alpha}$.} 
\textbf{a}-\textbf{c}, Three types of $\bm{\alpha}$-domain configurations that can be controlled by rotating the magnetic field in the $ab$-plane. The relative angle between the $a$-axis and $\bm{B}$ is $\phi$. The gray shaded wedges indicate the range of angles where a given domain is stable: $\phi=0^\circ \sim60^\circ$ ($180^\circ \sim 240^\circ$), $60^\circ \sim 120^\circ$ ($240^\circ \sim 300^\circ$), and $120^\circ \sim180^\circ$ ($300^\circ \sim 360^\circ$), respectively, for Configuration 1, 2, and 3. (Caption continues on next page.)
}
\label{main:AMR_domains}
\end{figure}

\addtocounter{figure}{-1}
\clearpage
\begin{figure}[h]%
\centering
\caption{\color{black}(Caption continues from previous page.)
\textbf{d}, Polar plot of AMR with illustrations of $\bm{\alpha}$-domain configurations. Step-like AMR observed in the $p$-wave phase originating from the switching of $\bm{\alpha}$-domains. 
For current $\bm{I}_a \parallel a$, a higher resistance $R^a_{xx}$ appears in Configuration 2 (red region), where $\bm{I}$ is nearly parallel to $\bm{\alpha}$. We also show the measurement configuration of anisotropic magnetoresistance (AMR). The current $I_a$ is applied along the $a$-axis, while the magnetic field $\bm{B}$ is rotated within the $ab$-plane. 
\textbf{e},\textbf{f}, AMR of a fabricated device with strain relief at $T = 2 \, \mathrm{K}$ and $30 \, \mathrm{K}$. The second $y$-axis on the right-hand side shows the relative magnitude of the AMR.
\textbf{g},\textbf{h}, AMR of a bulk sample at $T = 2 \, \mathrm{K}$ and $30 \, \mathrm{K}$. Bulk and device curves are in good qualitative agreement, although the AMR's relative magnitude is three times larger in the suspended device. We note that the AMR measurement on devices allows us to measure the resistivity along several current directions simultaneously (see Fig.~3, Main Text). 
\textbf{i},\textbf{j}, Separation of AMR into (i) step-type anomaly related to resistivity anisotropy in the $p$-wave magnet (green line, right axis) and (ii) six-fold patterns with the periodicity of the underlying crystal lattice (blue symbols, left axis). As the relative magnitude of the $p$-wave AMR is larger in the device, the six-fold patterns are harder to distinguish. Arrows in \textbf{e},\textbf{g} indicate the sweep direction of the magnetic field angle $\phi$.
}
\end{figure}

\clearpage
\begin{figure}[h]%
\centering
\includegraphics[width=1.0\textwidth]{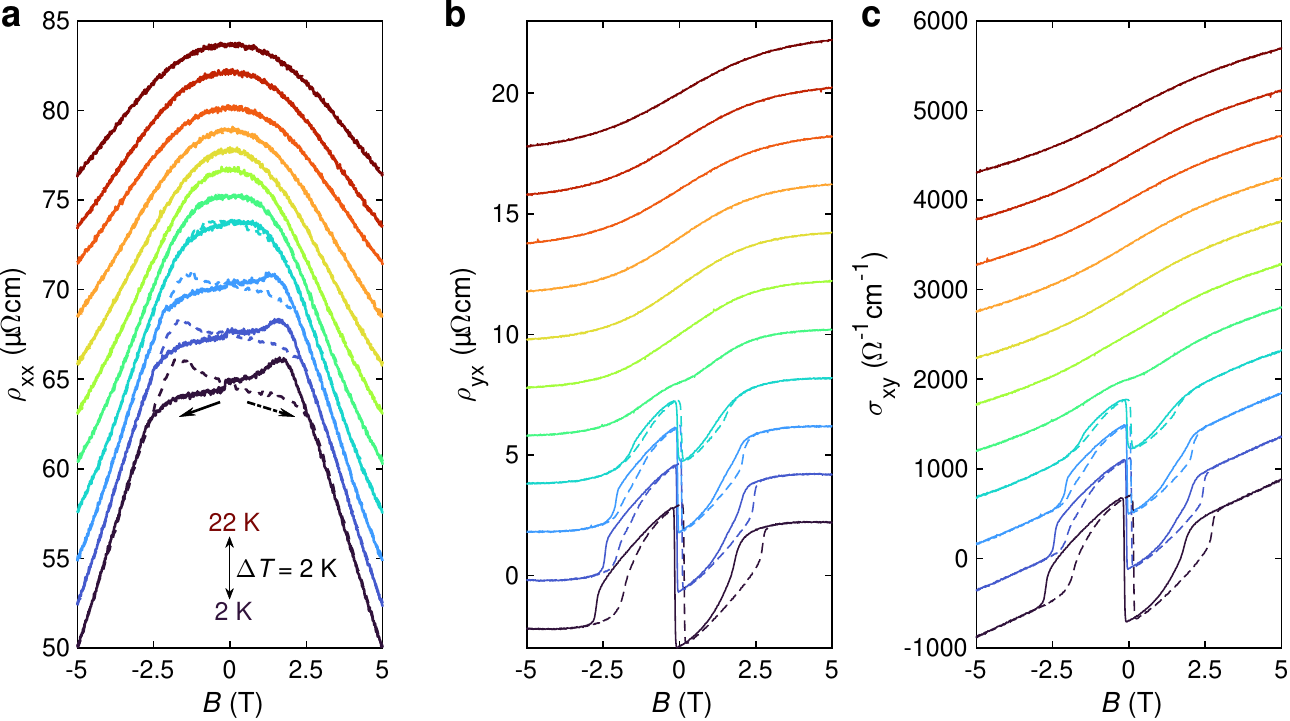}
\caption{\textbf{Electrical resistivity and Hall resistivity data for a bulk single crystal of \GRRAspec{}}.
\textbf{a}-\textbf{c}, Longitudinal resistivity $\rho_{xx}$, Hall resistivity $\rho_{yx}$, and Hall conductivity $\sigma_{xy}$ at various temperatures. In support of Fig.~\ref{main:Fig3} of the Main Text, here we show raw data of each quantity. Direction of magnetic field ramping is indicated by black arrows, with solid lines for {\color{black}$\partial B / \partial t<0$} and dashed lines for {\color{black}increasing} magnetic field. The magnetic field is applied along the $c$-axis of the crystal. The offset between curves is set to be $2 \, \mu\Omega\mathrm{cm}$, $2 \, \mu\Omega\mathrm{cm}$, and $500 \, \mathrm{\Omega}^{-1}\mathrm{cm}^{-1}$ in panels \textbf{a}, \textbf{b}, and \textbf{c}, respectively.
}
\label{main:Efig_Electric_transport}
\end{figure}

\clearpage
\begin{figure}[h]%
\centering
\includegraphics[width=1.0\textwidth]{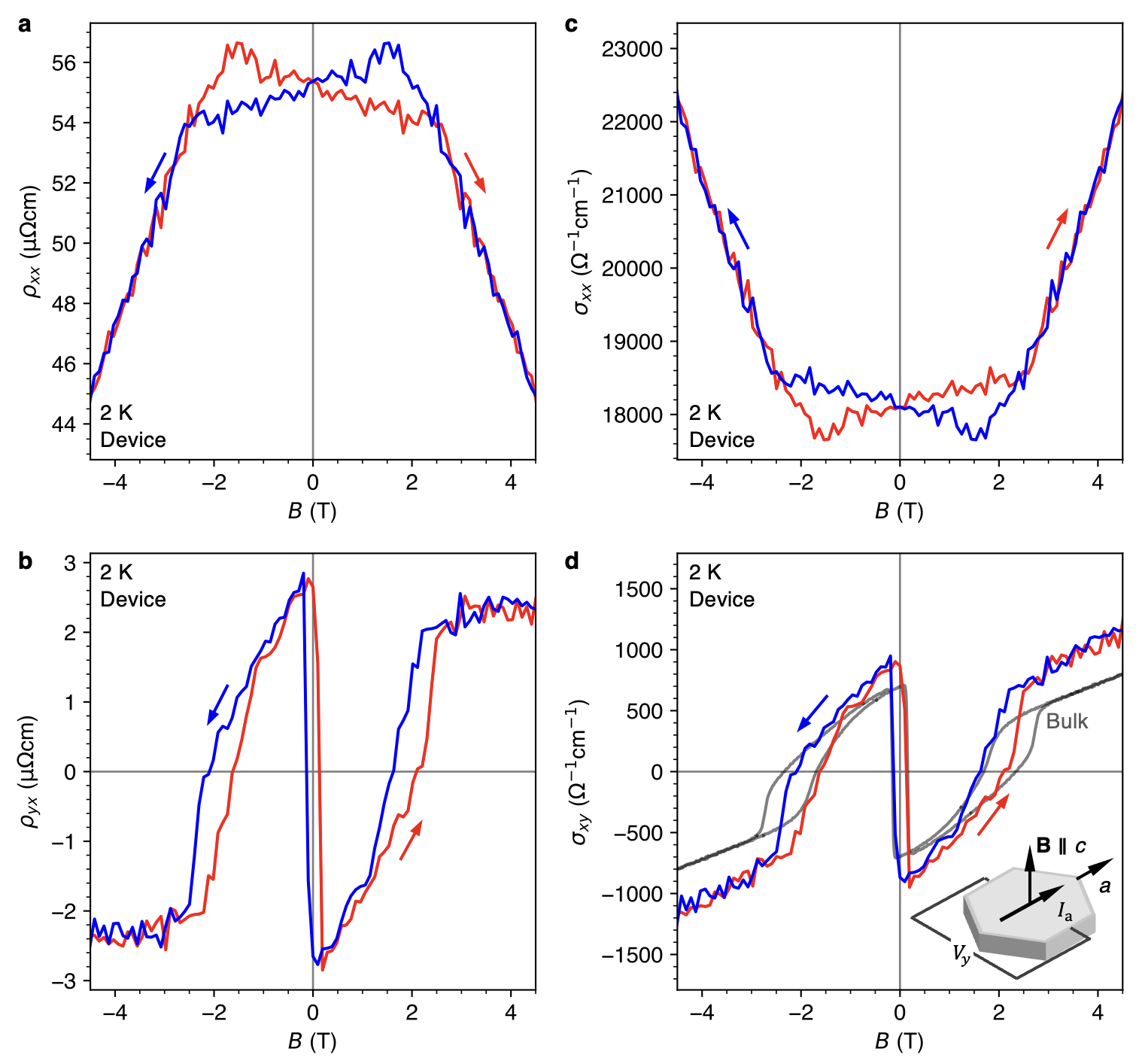}
\caption{\textbf{Comparing transport properties of bulk single crystal and suspended} device at $T = 2\,$K. The electric current and magnetic field are applied along the $a$-axis and $c$-axis, respectively. 
 \textbf{a} Longitudinal resistivity $\rho_{xx}$ for a device fabricated by the focused ion beam technique. \textbf{b}, \textbf{c} Magnetic field dependence of Hall resistivity $\rho_{yx}$ and longitudinal conductivity $\sigma_{xx}$. \textbf{d} Hall conductivity $\sigma_{xy}$ calculated for the device (red/blue) and for the bulk single crystal (gray line). The data are in good qualitative agreement. Inset: schematic measurement geometry. The hysteresis in $\rho_{xx}$ and $\sigma_{xx}$ is attributed to the texture of $p$-wave domains, as discussed further in Supplementary Fig.~\ref{main:RT_hysteresis}. To calculate the conductivities and resistivities of the device, we approximate the width and length of the central circular patch as a rectangular bar, implying significant systematic errors due to the geometry of the device.  Note: The Hall conductivity data presented in the Main Text was obtained from bulk single crystals to suppress systematic measurement errors for the absolute value of $\sigma_{xy}$. }
\label{main:SFig_device_Bdep}
\end{figure}

\clearpage
\begin{figure}[h]%
\centering
\includegraphics[width=1.0\textwidth]{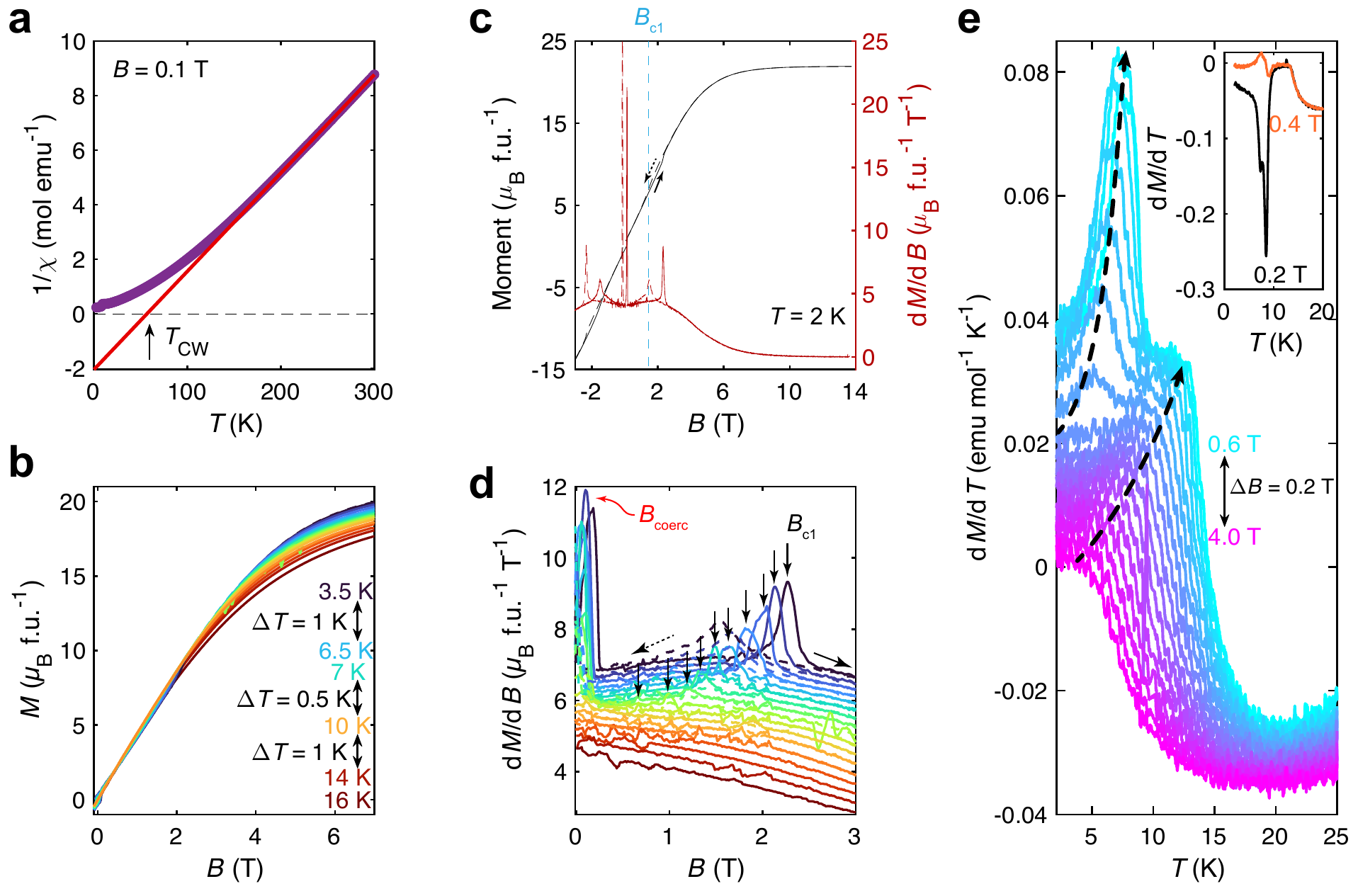}
\caption{\textbf{Bulk magnetization data with $\bm{B}\parallel c$}.
\textbf{a}, Curie-Weiss fit of magnetic susceptibility $ \chi = M/H$, measured with $0.1\,\mathrm{T}$ external magnetic field applied along the $c$-axis. \textbf{b}, $M(H)$ curves obtained at constant temperatures $T$ as indicated in the panel. \textbf{c}, Magnetization isotherm and its corresponding field derivative at $T=2 \, \mathrm{K}$. The dashed vertical line, labelled as $B_{\mathrm{c1}}$, represents the hysteretic boundary between Phase-I and Phase-II. Around this region, ramping up and ramping down direction of external magnetic field is indicated by solid and dashed arrow, respectively. \textbf{d}, Derivative of magnetization with respect to applied field at various $T$, showing peak-like features at  $B_\mathrm{coerc}$ (red arrow) and $B_{c1}$ (black arrows) defined in Fig.~\ref{main:Fig4}a. Consecutive curves are shifted by $0.2\,\mu_{\mathrm{B}}\, \mathrm{f.u.}^{-1}\,\mathrm{T}^{-1}$ for better visibility. {\color{black} Similar to Panel-c, solid and dashed lines denote the direction of magnetic field ramping.} \textbf{e}, Temperature dependent derivative $dM/dT$ of magnetization data measured at constant magnetic field, showing two clear transitions, $T_{\mathrm{N1}}$ and $T_{\mathrm{N2}}$ as defined in Fig.~\ref{main:Fig3}. Each consecutive curve is shifted by $0.002\,\mathrm{emu~mol}^{-1}\,\mathrm{K}^{-1}$ for better visibility. Inset: Zoom-in of data at low magnetic field.
}
\label{main:Efig_Magnetization}
\end{figure}

\clearpage
\begin{figure}[htb!]%
\centering
\includegraphics[width=1.\textwidth]{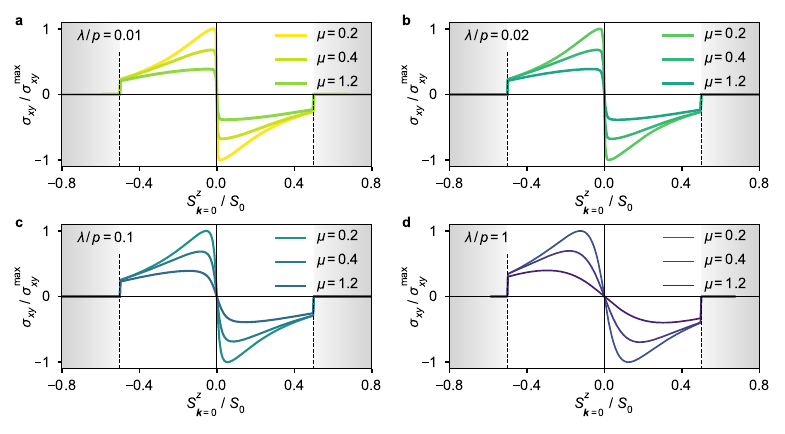}
\caption{
\textbf{Anomalous Hall conductivity of a distorted $p$-wave magnet.} 
Anomalous Hall conductivity as a function of the uniform net magnetization $S_{\bm{k}=0}^z$. Here, $S_0$ is the total length of the magnetic moment (saturation moment). In the low-energy model, only $m_z$ is varied while $p = 0.1$, $k_\mathrm{B}T = 0.04$ -- as well as spin-orbit coupling strength $\lambda$ and chemical potential $\mu$, written in each panel -- are fixed. We calculate $S_{\bm{k}=0}^z$ from $m_z/p$ and the parameters $J$, $t$ of a tight-binding Hamiltonian, as described in Supplementary Note~\ref{main:SI_AHC_calc}. A sharp anomaly emerges for small $S_{\bm{k}=0}^z$ when $\lambda$ is weak, $\lambda/p<1$. In the gray region, the spin helix collapses as discussed in Supplementary Note~\ref{main:conversion_exchange_splitting}.
}
\label{main:Hall_pwave_multipanel}
\end{figure}

\clearpage
\begin{figure}[h]%
\centering
\includegraphics[width=0.7\textwidth]{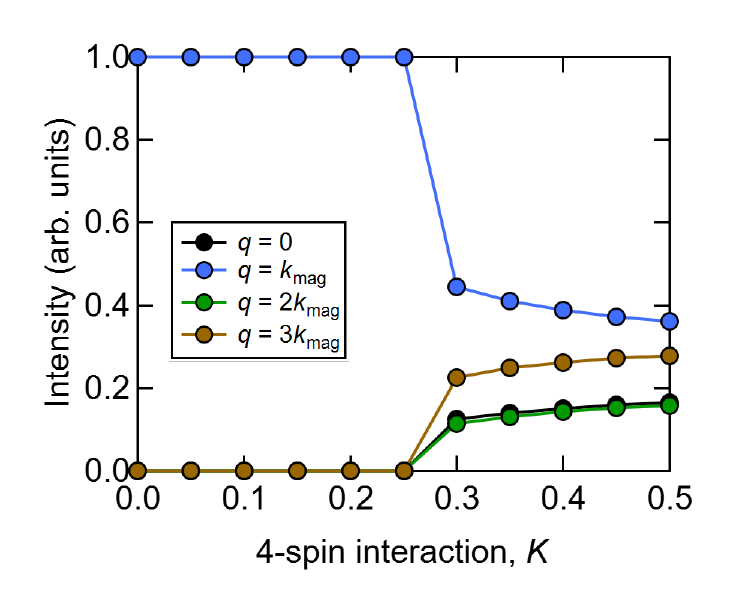}
\caption{\textbf{Solitonic spin states with a net magnetization from biquadratic magnetic interactions, following Supplementary Note \ref{main:SI_ModelDoubleq}}.
To explain the observed distorted spin helix state with a weak net magnetization, we develop a simple 1D two-parameter model in Eq.~(\ref{main:SpinModelDoubleQ}). In this model, we introduce the exchange anisotropy $\Delta > 0$ and $\Delta^{\prime} > 0$ to favour a spin helix with net magnetization along the $z$-axis; we also use the RKKY interaction $J = 1$, and the four spin interaction $K$. We optimize the energy to obtain the ground state of the model by simulated annealing. This figure shows the calculation results obtained for $\Delta = 0.1$. Magnetic states with higher harmonics and spontaneous magnetization appear when the four-spin interaction is sufficiently large, $K > 0.25$. 
}
\label{main:SolitonicSpins}
\end{figure}

\clearpage

\clearpage
\begin{table}	
    \centering
    \begin{tabular}{|c|c|c|c|c|c|c|c|}\hline
         \multirow{2}{*}{Material} & \multirow{2}{*}{Spin texture} & $\bm{q}$-vector & $B_\mathrm{sat}^\mathrm{est}$ & $\sigma_{xy}^\mathrm{A}$ & $\sigma_{xy}^\mathrm{A} / \sigma_{xx}$  & $T_\mathrm{max}$ & \multirow{2}{*}{Ref.} \\  
        & & (r.l.u.) & (T) & ($\mathrm{\Omega}^{-1}\mathrm{cm}^{-1}$) &(\%) & (K) &  \\ \hline \hline 
        \GRRAspec{} & non-collinear & $(\frac{1}{6},\frac{1}{6},0)$ & $5.5$ & $600$ & $4.2\, \mathchar`- \,4.7$ & $2$ & This work \\ \hline 
        $\mathrm{Mn}_{3}\mathrm{Sn}$ & non-collinear & $(0, 0, 0)$ & $660$ & $20\, \mathchar`- \,140$ & $0.4 \, \mathchar`- \, 2.4$ & $2$ & \cite{Nakatsuji2015, Ikhlas2017} \\ \hline 
        $\mathrm{Mn}_{3}\mathrm{Ge}$ & non-collinear & $(0, 0, 0)$ & $1500$ & $15 \, \mathchar`- \, 350$ & $0.4 \, \mathchar`- \, 3.6$ & $2$ &  \cite{Nayak2016, Chen2021}\\ \hline
        $\mathrm{Mn}_{3}\mathrm{Ga}$ & non-collinear & $(0, 0, 0)$ & $180$ & $18$ & $0.5$ & $300$ &  \cite{Liu2017}\\ \hline
        $\mathrm{Mn}_{3}\mathrm{Sb}$ & non-collinear & $(0, 0, 0)$ & $650$ & $310$ & $1.1\, \mathchar`- \, 1.5$ & $150$ &  \cite{Hayashi2023}\\ \hline
        $\mathrm{Mn}_{3}\mathrm{Pt}$ & non-collinear & $(0, 0, 0)$ & $2445$ & $100$ & $0.8$ & $120$ &  \cite{Cespedes2023}\\ \hline
        $\mathrm{CoNb}_{3}\mathrm{S}_{6}$ & non-coplanar & $(\frac{1}{2}, 0, 0)$ & $250$ & $27$ & $0.7$ & $23$ & \cite{Ghimire2018} \\ \hline
        $\mathrm{CoTa}_{3}\mathrm{S}_{6}$ & non-coplanar & $(\frac{1}{2}, 0, 0)$ & $170$ & $75 \, \mathchar`- \, 135$ & $0.9 \, \mathchar`- \, 1.6$ & $6$ & \cite{Park2023, Takagi2023} \\ \hline
        $\mathrm{Mn}_{5}\mathrm{Si}_{3}$ & non-coplanar & $(\frac{1}{2}, 0, 0)$ & $180$ & $160$ & $2.3$ & $25$ & \cite{Surgers2024} \\ \hline 
        NbMnP & non-collinear & $(0, 0, 0)$ & $1100$ & $230$ & $1.9 \, \mathchar`- \, 3.2$ & $2$ & \cite{Kotegawa2023} \\ \hline 
        TaMnP & non-ollinear & $(0, 0, 0)$ & $1200$ & $370$ & $1.4 \, \mathchar`- \, 2.7$ & $2$ & \cite{Kotegawa2024} \\ \hline 
        $\mathrm{Ce}_{2}\mathrm{CuGe}_{6}$ & collinear & $(0, 0, 0)$ & $90$ & $550$ & $0.2$ & $1.3$ & \cite{kotegawa2024PRL} \\ \hline 
        
    \end{tabular}
    \caption{\textbf{Comparison of anomalous Hall conductivity $\sigma_{xy}^\mathrm{A}$ in zero-field and saturation magnetic field $B_\mathrm{sat}^\mathrm{est}$ among bulk single crystalline magnets with small net magnetization.} \GRRAspec{} shows a commensurate non-collinear spin texture, a giant anomalous Hall conductivity, and a large Hall angle. It also has a low magnetic field for transition into the ferromagnetic state, simplifying the data analysis. $T_\mathrm{max}$ is a temperature where $\sigma_{xy}^\mathrm{A}$ takes a maximum value. Error bar of $\sigma_{xy}^\mathrm{A}/\sigma_{xx}$ in our material is estimated from the variation of the values between a bulk sample and a device. As for Mn$_3$Sn and Mn$_3$Ge, error bars for $\sigma_{xy}^\mathrm{A}$ and $\sigma_{xy}^\mathrm{A}/\sigma_{xx}$ indicate variation of the values for different field directions. Since $\sigma_{xy}^\mathrm{A}/\sigma_{xx}$ of Mn$_3$Sb, NbMnP, and TaMnP takes a maximum at temperatures different from $T_\mathrm{max}$ of $\sigma_{xy}^\mathrm{A}$, we added error bars for these compounds. As for CoTa$_3$S$_6$, error bars are added because there is a variation in reported values of $\sigma_{xy}^\mathrm{A}$. In Fig.~\ref{main:Fig4}e, we use $\sigma_{xy}^\mathrm{A}$ and $\sigma_{xy}^\mathrm{A}/\sigma_{xx}$ at $T_\mathrm{max}$, and also add error bars for $\sigma_{xy}^\mathrm{A}/\sigma_{xx}$ based on this table.\\
    }
    \label{main:Table_AHC}
\end{table}

\clearpage

\end{document}